\documentclass[journal]{IEEEtran}

\usepackage{float}
\usepackage{pgfplots}
\pgfplotsset{compat=1.18}

\newcommand{\figheight}{0.7\linewidth}
 
\usepackage{KalmanNet} 
 \usepackage[all=normal,paragraphs=tight,floats=normal,mathspacing=normal,wordspacing=tight,charwidths=tight,mathdisplays=normal,leading=normal]{savetrees}

\usepackage{xcolor}

\usepackage{amsmath,amssymb,mathtools}
\usepackage{bm}
\let\oldnl\nl
\newcommand{\nonl}{\renewcommand{\nl}{\let\nl\oldnl}}

\IEEEoverridecommandlockouts

\begin{document}

\title{EM-KalmanNet: Learned Expectation-Maximization for Adaptive Tracking in Partially Known, Block-Wise Time-Varying State-Space Models}

\author{
	\IEEEauthorblockN{Ori Cohen, Nir Shlezinger,  \IEEEmembership{Senior Member, IEEE}, and Tirza Routtenberg,  \IEEEmembership{Senior Member, IEEE}
\thanks{Parts of this work were presented at the 2026 Sensor Array and Multichannel Signal Processing Workshop (SAM) as the paper~\cite{Cohen2026EMKalmanNet}.
 O. Cohen, N. Shlezinger, and T. Routtenberg are with the School of ECE, Ben-Gurion University of the Negev, Israel (e-mail: orico@post.bgu.ac.il; \{nirshl; tirzar\}@bgu.ac.il). \\
  This work was supported by the Israel Science Foundation (ISF) under grant no. 3314/25 and grant no. 1148/22. 
}}
}

\maketitle


\begin{abstract}
State estimation in partially known state space (SS) models is challenging when the dynamics or observation model varies across short data blocks. Classical model-based approaches, such as the expectation-maximization (EM) Kalman filter, jointly recover the latent states and the unknown model parameters, but rely on linear-Gaussian assumptions that should accurately describe the system and require numerous forward--backward passes. Consequently, their performance and computational efficiency may deteriorate under complex and non-stationary real-world conditions. On the other hand, learned Kalman smoothers are robust to model mismatch yet cannot adapt at inference to unseen model variations without labeled data. 
In this work, we propose \emph{EM-KalmanNet}, an AI-aided tracking algorithm for adaptive smoothing in block-wise time-varying SS models.
The method unfolds a fixed, small number of EM-like iterations into a trainable architecture: a parameter-aware RTSNet implements a learned E-step conditioned on the current model-parameter estimate, while a lightweight M-Net implements a learned M-step that updates the state-transition or the observation matrix using empirical moments, residuals, and gradient-related statistics. The two modules are shared across the unfolded iterations and are trained offline via a dedicated three-stage procedure. During deployment, the per-block parameter estimate is propagated between consecutive blocks, enabling observation-driven adaptation without labeled online data or knowledge of the noise statistics. Experiments involving linear and nonlinear models, Gaussian and non-Gaussian noise, Lorenz attractor tracking, and acoustic source localization demonstrate that EM-KalmanNet consistently outperforms model-based and data-driven benchmarks while substantially reducing inference latency relative to the EM-KF.
\end{abstract}


\acresetall

\section{Introduction}
\label{sec:intro} 
Tracking the latent states of dynamical systems from noisy observations is a fundamental problem in signal processing~\cite{sarkka2023bayesian}, with applications ranging from navigation and localization to smart transportation and autonomous systems~\cite{karami2020smart}. Classical tracking methods are effective when the statistical relations governing the state evolution and observations are accurately known. Often in practice, however, these relations are only partially characterized and may themselves evolve over time as the operating conditions, motion patterns, or sensing environment change~\cite{bar2004estimation}. When such variations are known, they can be directly incorporated into a time-varying \ac{ss} model~\cite{durbin2012time}; the more challenging setting arises when the variations are unknown and must be inferred from the same noisy observations used for state estimation. This difficulty is further amplified when changes occur rapidly, such that only short observation intervals are available for identifying the current dynamics.

Model-based tracking is commonly built upon the \ac{kf}~\cite{kalman1960new}, the \ac{rts} smoother~\cite{rauch1965maximum}, and their nonlinear extensions~\cite{schmidt1981kalman,julier2004unscented,arasaratnam2009cubature}. When the governing model is unknown or time varying, these estimators can be augmented with mechanisms for system identification and parameter adaptation~\cite{mehra1970identification}. Representative approaches include multiple-model tracking for switching dynamics~\cite{blom1988interacting}, joint or dual state and parameter estimation~\cite{wan2001dual,Togneri2003}, and adaptive estimation of unknown noise statistics~\cite{zhang2020identification}. 
A particularly principled approach is the \ac{em} algorithm~\cite{moon1996expectation}, which enables joint inference of latent states and unknown model parameters by alternating between smoothing under the current parameter estimate and updating the parameters from the inferred state statistics~\cite{gannot2008kalman,Shumway1982}. However, such model-based adaptive methods remain tied to the assumed probabilistic description: their reliability can deteriorate under model mismatch and non-Gaussian disturbances, while standard closed-form \ac{em}-\ac{kf} updates are primarily associated with linear-Gaussian \ac{ss} models. Moreover, iterative parameter estimation methods such as \ac{em} require repeated forward-backward passes and typically benefit from sufficiently long observation sequences, making reliable low-latency adaptation particularly challenging when the dynamics change over short time intervals.

Data-driven state estimators provide an alternative by learning aspects of the underlying dynamics and inference rule directly from data~\cite{shlezinger2025artificial}. Deep latent \ac{ss} models and recurrent estimators have been developed to learn nonlinear temporal dependencies and state representations without requiring a complete analytical model~\cite{haarnoja2016backprop,becker2019recurrent,ghosh2023danse,satorras2019combining,klushyn2021latent}. 
A complementary line of work follows the model-based deep learning paradigm and retains the algorithmic structure of classical Bayesian estimators to leverage partial characterization of the dynamics while learning those components that are difficult to specify analytically~\cite{shlezinger2020model}. For example, EKFNet~\cite{Xu2024EKFNet} differentiates through the EKF to learn unknown noise covariance parameters from data while retaining the model-based filtering structure. In particular, KalmanNet~\cite{revach2022kalmannet} replaces the analytically computed Kalman gain with a learned recurrent mapping, allowing \ac{kf}-type tracking under partially known dynamics and unknown noise statistics, while RTSNet~\cite{ni2022rtsnet} extends this methodology to smoothing. Related variants further improve robustness or accommodate high-dimensional observations, and alternative learned gain architectures~\cite{choi2023split,buchnik2023latent,wang2024nonlinear}. These approaches can substantially reduce sensitivity to inaccuracies in the assumed \ac{ss} model; nonetheless, their learned inference rules are determined by the distributions represented during training, and can consequently deteriorate when the model encountered at deployment deviates from the training conditions.

Several approaches have therefore sought to endow learned Kalman-type estimators with adaptation capabilities. Unsupervised KalmanNet~\cite{revach2022unsupervised} exploits observation-prediction errors to adapt a pre-trained KalmanNet to changing \ac{ss} models without requiring ground-truth state labels. However, it requires updating the learned model through lengthy gradient-based optimization at deployment. While such gradient-based learning can be made more efficient by drift-aware learning rate adaptation~\cite{zhang2026change} or by incorporating model-agnostic meta-learning techniques~\cite{chen2025maml}, such a training method may still be unsuitable for coping with rapid variations. Adaptive KalmanNet~\cite{ni2024adaptive} uses a compact hypernetwork to modulate the learned filtering rule according to context information describing the current operating regime, enabling adaptation without retraining, assuming that the variations are encompassed by the offline training data, and that such contextual information is available.   
Still, there remains a gap between classical adaptive estimators, which can infer changing model parameters directly from observations but rely strongly on prescribed statistical models, and learned state estimators, which are robust to model mismatch but lack an efficient mechanism for jointly inferring the latent states and unknown time-varying model parameters from unlabeled observations.

 Motivated by this gap, we propose {\em EM-KalmanNet}, a model-based deep learning methodology for adaptive smoothing in partially known, block-wise time-varying \ac{ss} models. EM-KalmanNet combines the observation-driven parameter adaptation of the \ac{em}-\ac{kf} with the robustness of KalmanNet-type learned estimators by unfolding a fixed, small number of \ac{em}-like iterations into a discriminative trainable architecture~\cite{shlezinger2022discriminative}. Rather than retaining the analytical \ac{em} updates, we learn both stages of the iterative procedure: a parameter-aware RTSNet~\cite{ni2022rtsnet} implements the smoothing step conditioned on the current model-parameter estimate, while a lightweight module termed {\em M-Net} learns to refine this estimate from model-inspired statistics of the smoothed trajectory. The resulting architecture is trained entirely offline and, at deployment, jointly refines the latent states and the unknown model parameters directly from the observations, without labeled online data, online network optimization, or knowledge of the noise statistics. 
 
 Our main contributions are summarized as follows:
\begin{itemize}
\item We introduce an \ac{em}-inspired neural architecture for adaptive smoothing in partially known and time-varying \ac{ss} models. The proposed EM-KalmanNet unfolds a prescribed number of \ac{em}-like iterations and replaces their analytical components with shared trainable modules, yielding a fixed-complexity inference procedure that combines model-based structure with data-driven robustness. We consider both unknown time-varying state-transition and observation models, while allowing the available part of the \ac{ss} model to be nonlinear and the noise statistics to be unknown.
\item We design dedicated learned E- and M-steps for joint state and parameter inference. The E-step is implemented by a parameter-aware RTSNet whose model-based recursions and learned gains are conditioned on the current parameter estimate. The M-step is realized by an M-Net that directly learns a correction to the unknown model parameter from compact empirical moments, residual statistics, and gradient-related quantities motivated by the analytical \ac{em} update. We further show that a covariance-agnostic counterpart of the classical \ac{em} M-step can be expressed within this correction-based formulation, while the learned M-Net is free to realize more general updates suitable for model mismatch and non-Gaussian settings.
\item We develop a dedicated multi-stage training procedure for the unfolded architecture. The parameter-aware RTSNet is first trained to smooth under both accurate and perturbed parameter values, after which the M-Net is trained to perform parameter refinement with the smoother fixed, followed by joint fine-tuning of the complete unfolded estimator. Sharing the learned modules across iterations and training them to operate on imperfect incoming parameter estimates enables observation-driven adaptation with a small prescribed inference budget and permits the parameter estimate to be propagated between consecutive blocks.
\item We evaluate EM-KalmanNet across a broad set of time-varying tracking scenarios involving unknown state-transition and observation models, nonlinear dynamics and observations, Gaussian and non-Gaussian disturbances, short observation blocks, Lorenz-system tracking, and acoustic source localization. The results demonstrate robust adaptation under model mismatch and changing dynamics, with improved state-estimation performance over mismatched model-based and learned benchmarks across the considered settings, while requiring substantially lower inference latency than iterative \ac{em}-\ac{kf} processing. 

\end{itemize}

The rest of this paper is organized as follows: Section~\ref{sec:system-model-and-preliminaries} formulates the problem. EM-KalmanNet is detailed in Section~\ref{sec:method}, and evaluated in Section~\ref{sec:numerical_study}.  Section~\ref{sec:conc} provides concluding remarks.

\smallskip
\noindent\textbf{Notation:}
Vectors and matrices are denoted by  bold lowercase and uppercase letters, respectively. The operators $(\cdot)^{\top}$, $\operatorname{vec}(\cdot)$, $\mathbb{E}[\cdot]$, and $\operatorname{Cov}(\cdot)$ denote transpose, vectorization, expectation, and covariance, respectively.
Moreover, $\|\cdot\|_2$ and $\|\cdot\|_{\mathrm F}$ denote the Euclidean and Frobenius norms, and $\mathbf{I}_m$ is the $m\times m$ identity matrix.
For a sequence $\{\myVec{a}_t\}_{t=1}^{T}$, we use
$\myVec{a}_{1:T}\triangleq(\myVec{a}_1,\ldots,\myVec{a}_T)$
to denote the corresponding length-$T$ sequence.

\section{System Model and Preliminaries}
\label{sec:system-model-and-preliminaries} 
In this section, we present the system model and the required background for the proposed architecture.
 In Subsection~\ref{ssec:SS}, we formulate the block-wise time-varying \ac{ss} model and present the main challenges. In  Subsection~\ref{ssec:Preliminaries}, we review the main building blocks underlying EM-KalmanNet, including Kalman smoothing, EM-KF procedures, and RTSNet, thereby setting the basis for the EM-KalmanNet architecture developed in Section~\ref{sec:method}.

\subsection{System Model} 
\label{ssec:SS}
{\bf Signal Model:} 
We consider a discrete-time dynamic system with hidden state $\myVec{x}_t \in \mathbb{R}^{m}$ that is related to an observation $\myVec{y}_t \in \mathbb{R}^{n}$ according to a \ac{ss} model of the form
\begin{align}
\label{eqn:ssModel}
    \myVec{x}_t &= f_t(\myVec{x}_{t-1}) + \myVec{w}_t, \quad 
    \myVec{y}_t = h_t(\myVec{x}_t) + \myVec{v}_t.
\end{align}
In \eqref{eqn:ssModel}, $f_t(\cdot)$ is the state evolution function, and $h_t(\cdot)$ is the observation function. The process and observation noises,  $\myVec{w}_t$ and $\myVec{v}_t$, respectively, are assumed to be zero-mean,
mutually
independent, temporally independent, and independent of the initial state, $  \myVec{x}_0$. 
We focus on settings in which the underlying \ac{ss} model varies over time in a block-wise manner, such that the state-transition and observation functions, $f_t(\cdot)$ and $h_t(\cdot)$, remain fixed within each block and may vary between consecutive blocks. Each block consists of $T$ time instants.
For a block containing times $\tau+1,\ldots,\tau+T$, $\mathbf{x}_{\tau}$ serves as its initial state for the corresponding \ac{ss} model. 

\smallskip
{\bf Problem Formulation:}
We consider a block-wise smoothing problem in which, at block index $i$, 
$\myMat{X}_i = \{\myVec{x}_\tau\}_{\tau = (i-1)\cdot T+1}^{i\cdot T}$ 
is estimated from the observed sequence 
$\myMat{Y}_i = \{\myVec{y}_\tau\}_{\tau = (i-1)\cdot T+1}^{i\cdot T}$. 
Each trajectory consists of $M$ consecutive blocks, 
where the \ac{ss} model is fixed within each block and 
may vary between blocks.
Given $\myMat{Y}_i$, the prior first- and second-order moments of the initial state $\myVec{x}_{(i-1)T}$ obtained from the previous block, and an initial estimate of the unknown parameter, the goal is to estimate $\myMat{X}_i$ while adapting the unknown parameter.

We particularly focus on smoothing subject to the following challenges:
\begin{enumerate}[label=\textbf{C.\arabic*}, leftmargin=*, itemsep=0pt]
    \item \label{itm:Approx}  The functions $f_t(\cdot)$ and $h_t(\cdot)$ may be crude approximations of the true underlying model.
    
    \item \label{itm:Noise} The distribution of the noise signals is unknown and may be non-Gaussian.
    
    \item \label{itm:varying} The variations in $f_t(\cdot)$ and $h_t(\cdot)$ are unknown.

    \item \label{itm:latency} The estimate of $\myMat{X}_i$ must be produced with minimal latency. 
    \item \label{itm:shortBlocks}
    Variations in $f_t(\cdot)$ or $h_t(\cdot)$ can occur rapidly, requiring operation with a small block length $T$, which leaves few observations for reliable smoothing and parameter adaptation.
\end{enumerate}
To cope with \ref{itm:Approx}-\ref{itm:shortBlocks}, we assume that during training, one has access to data from the underlying \ac{ss} model. Such data are given by
\begin{equation}
\label{eq:dataset}
\mySet{D}
= \Big\{  \Big\{\{(\myVec{x}_{j,i,t}, \myVec{y}_{j,i,t})\}_{t=1}^{T}, 
f_{j,i}, h_{j,i}\Big\}_{i=1}^{M}
\Big\}_{j=1}^{N_{\mathcal D}}.
\end{equation}
Here, $N_{\mathcal D}$ denotes the number of training trajectories, while$f_{j,i}$ and $h_{j,i}$ denote the nominal state-transition and observation models available to the estimator during training. Such side information naturally arises in model-based signal processing systems from first-principles modeling, system identification, or calibration, and may itself be approximate, as also demonstrated in the experimental study.

\smallskip
{\bf Cases Considered:}
Throughout the paper, we focus on two complementary scenarios, distinguished by which of the two functions in \eqref{eqn:ssModel} is unknown.  In both cases, the available function is used as is, while the unknown function is estimated from data and is parameterized as linear within each block. \\
{\bf Case 1 (Unknown State Transition Function):}
The state-transition function $f_t(\cdot)$ is unknown and varies between blocks; it is parameterized by a transition matrix $\myMat{F}$, while $h_t(\cdot)$ is available. In this case, \eqref{eqn:ssModel} is represented, within a given block, by the model
\begin{align}
\label{eqn:ssModelSurF}
\myVec{x}_t &= \myMat{F}\myVec{x}_{t-1} + {\myVec{w}}_t, \quad
\myVec{y}_t = h_t(\myVec{x}_t) + {\myVec{v}}_t.
\end{align}
{\bf Case 2 (Unknown Observation Function):}
The observation function $h_t(\cdot)$ is unknown and varies between blocks; it is parameterized by an observation matrix $\myMat{H}$, whereas the state-transition function $f_t(\cdot)$ is available. In this case, \eqref{eqn:ssModel} is represented, within a given block, by the model
\begin{align}
\label{eqn:ssModelSurH}
\myVec{x}_t &= f_t(\myVec{x}_{t-1}) + {\myVec{w}}_t, \quad
\myVec{y}_t = \myMat{H}\myVec{x}_t + {\myVec{v}}_t.
\end{align}

\subsection{Preliminaries}
\label{ssec:Preliminaries}

{\bf Kalman Smoothing:} 
The classical approach for smoothing is based on variants of the \ac{kf}, such as the extended \ac{rts} algorithm~\cite{rauch1965maximum}.
For linear Gaussian \ac{ss} models with known parameters, the \ac{kf} is optimal for online filtering, while the \ac{rts} smoother is optimal for block smoothing in the \ac{mmse} sense.
Such methods are designed for dynamics that are captured by a known \ac{ss} model (violating \ref{itm:Approx} and \ref{itm:varying}) with known Gaussian state and observation noise covariances $\myMat{Q}$ and $\myMat{R}$, respectively (violating~\ref{itm:Noise}).

To describe the extended \ac{rts} smoother, we focus on  a single block of length $T$, indexed locally as $t=1,\ldots,T$. The initial prior information is represented by ${\myVec{x}}_{0}$, from which the state trajectory evolves according to the \ac{ss} model. The smoother  applies a forward pass followed by a backward pass to estimate the latent state trajectory $\myVec{x}_{1:T}$ within the block. 
The forward pass is an \ac{ekf}, which updates its \textit{prior} state and covariance estimates based on past observations. For each $t$, the forward pass computes the predicted first-order moments via
\begin{equation}
\hat{\myVec{x}}_{t\given{t-1}}=f_t\brackets{\hat{\myVec{x}}_{t-1|t-1}}, \quad \hat{\myVec{y}}_{t\given{t-1}} =
h_t\brackets{\hat{\myVec{x}}_{t\given{t-1}}},    
\end{equation}
and the predicted second-order moments via
\begin{equation*}
\hat\mySigma_{t\given{t-1}} =\hat{\myMat{F}}_t\hat\mySigma_{t-1\given{t-1}}\hat{\myMat{F}}_t^\top+\myMat{Q}, ~ \hat{\myMat{S}}_{t} =\hat{\myMat{H}}_t\hat\mySigma_{t\given{t-1}}\hat{\myMat{H}}_t^\top+\myMat{R},     
\end{equation*}
where  $
\hat{\myMat{F}}_t
=
\left.
\tfrac{\partial f_t(\myVec{x})}
{\partial\myVec{x}}
\right|_{\myVec{x}=\hat{\myVec{x}}_{t-1|t-1}}$, and 
$\hat{\myMat{H}}_t
=
\left.
\tfrac{\partial h_t(\myVec{x})}
{\partial\myVec{x}}
\right|_{\myVec{x}=\hat{\myVec{x}}_{t|t-1}}$ 
denote the  local linearizations of $f_t(\cdot)$ and $h_t(\cdot)$.
 The predictions are updated via
 \begin{equation}
 \label{eq:FW_update_1}    \hat{\myVec{x}}_{t\given{t}}=\hat{\myVec{x}}_{t\given{t\!-\!1}}\!+\!\Kgain_{t}(\myVec{y}_t\!-\!\hat{\myVec{y}}_{t\given{t\!-\!1}}), \,  \hat{\mySigma}_{t\given{t}}= \hat{\mySigma}_{t\given{t\!-\!1}}\!-\!\Kgain_{t}\hat{\myMat{S}}_{t}\Kgain^{\top}_{t},  
 \end{equation}
 with 
 \emph{forward} \ac{kg}  $\Kgain_{t}=\hat{\mySigma}_{t\given{t-1}}{\hat{\myMat{H}}}_t^\top\hat{\myMat{S}}^{-1}_{t}$.

The backward pass refines the forward estimates by incorporating information from future observations, applied sequentially from $t=T-1$ to $t=1$. For each $t$, the state is updated via 
\begin{align}\label{eq:BW_update_1}
\hat{\myVec{x}}_{t}&=
\hat{\myVec{x}}_{t\given{t}}+\Sgain_t \big(\hat{\myVec{x}}_{t+1} -  {\hat{\myVec{x}}_{t+1\given{t}}}\big),
\end{align}
while the corresponding covariance is updated via
\begin{align}
\hat{\mySigma}_{t}=\hat{\mySigma}_{t|t} + 
\Sgain_t(\hat{\mySigma}_{t+1}-\hat{\mySigma}_{t+1\given{t}})\Sgain^{\top}_{t}.
\end{align}
Here, the backward \ac{kg} is $\Sgain_t=\hat{\mySigma}_{t|t}  \hat{\myMat{F}}_{t+1}^{\top}\hat{\mySigma}_{t+1\given{t}}^{-1}$. 


\smallskip
{\bf EM-KF:} 
The classical treatment of smoothing with unknown parameters considers a linear Gaussian instance of \eqref{eqn:ssModel}, i.e., $f_t(\myVec{x})=\myMat{F}\myVec{x}$,  $h_t(\myVec{x})=\myMat{H}\myVec{x}$, and Gaussian ${\myVec{w}}_t$ and ${\myVec{v}}_t$ with known covariances $\myMat{Q},\myMat{R}$. 
 In this setting, the \ac{em} algorithm can  estimate the unknown parameters and the latent states~\cite{Shumway1982}.

The \ac{em} algorithm iteratively forms the expected complete-data log-likelihood of the observations,
\begin{align}
\mathcal{Q}(\boldsymbol{\phi};\boldsymbol{\phi}^{(l-1)})
=
\mathbb{E}_{\myVec{x}_{1:T}|\myVec{y}_{1:T};\boldsymbol{\phi}^{(l-1)}}
\left[
\log p(\myVec{x}_{1:T},\myVec{y}_{1:T};\boldsymbol{\phi})
\right],
\label{eqn:EMform}
\end{align}
and then selects the parameter that maximizes this quantity.
In \eqref{eqn:EMform}, the unknown parameter is $\boldsymbol{\phi}$, and $\hat{\boldsymbol{\phi}}^{(l)}$ is its estimate after $l$ EM iterations. The method starts with an initial estimate $\boldsymbol{\phi}^{(0)}$,   and alternates between two steps:\\
$(i)$ The {\em E-step} applies the \ac{rts} smoother under the current parameter estimate to compute the posterior moments of the latent state trajectory required for the expected complete-data log-likelihood. At iteration $l$, it assumes a linear Gaussian \ac{ss} model with the current parameter estimate $\boldsymbol{\phi}^{(l-1)}$, and computes
\[
\left\{
\hat{\myVec{x}}_{t}^{(l)},
\hat{\mySigma}_{t}^{(l)},
\hat{\mySigma}_{t,t-1}^{(l)}
\right\}_{t=1}^{T},
\]
where the lag-one smoothed cross-covariance is given by
\begin{equation}
    \hat{\myMat{P}}_{t,t-1}^{(l)}=\mathrm{Cov}\big(\myVec{x}_t,\myVec{x}_{t-1}|\myVec{y}_{1:T} ; \boldsymbol{\phi}^{(l-1)} \big) ,     
\end{equation}
 computed using the standard RTS smoother recursion.

$(ii)$ The \emph{M-step} updates the unknown parameter by maximizing the expected log-likelihood using the moments computed in the E-step.
To simplify the M-step notation, we define the following EM sufficient statistics:
\begin{subequations}
    \label{eqn:A_def}
\begin{align}
\label{A1_def}
\myMat{A}_1^{(l)}
&\triangleq
\sum_{t=1}^{T}
\left(
\hat{\myVec{x}}_{t}^{(l)}
(\hat{\myVec{x}}_{t-1}^{(l)})^\top
+
\hat{\mySigma}_{t,t-1}^{(l)}
\right), \\
\label{A2_def}
\myMat{A}_2^{(l)}
&\triangleq
\sum_{t=1}^{T}
\left(
\hat{\myVec{x}}_{t-1}^{(l)}
(\hat{\myVec{x}}_{t-1}^{(l)})^\top
+
\hat{\mySigma}_{t-1}^{(l)}
\right), \\
\label{A3_def}
\myMat{A}_3^{(l)}
&\triangleq
\sum_{t=1}^{T}
\left(
\hat{\myVec{x}}_{t}^{(l)}
(\hat{\myVec{x}}_{t}^{(l)})^\top
+
\hat{\mySigma}_{t}^{(l)}
\right), \\
\label{A4_def}
\myMat{A}_4^{(l)}
&\triangleq
\sum_{t=1}^{T}
\myVec{y}_{t}
(\hat{\myVec{x}}_{t}^{(l)})^\top .
\end{align}
\end{subequations}
The sufficient statistics in \eqref{eqn:A_def}
also depend on the initial smoothed moments
$\hat{\myVec{x}}_{0}^{(l)}$,
$\hat{\myMat{P}}_{0}^{(l)}$,
$\hat{\myMat{P}}_{1,0}^{(l)}$,
which are obtained from the initialization of the RTS smoother.

When the state-transition matrix $\myMat{F}$ is unknown, the M-step yields
\begin{align}
\myMat{F}^{(l)}
=
\myMat{A}_1^{(l)}
\left(
\myMat{A}_2^{(l)}+\epsilon \cdot \myMat{I}_m
\right)^{-1},
\label{eqn:MStepF}
\end{align}
where $\epsilon > 0$ is a small regularization constant.
When the observation matrix $\myMat{H}$ is unknown, the M-step yields
\begin{align}
\myMat{H}^{(l)}
=
\myMat{A}_4^{(l)}
\left(
\myMat{A}_3^{(l)}+\epsilon \cdot \myMat{I}_m
\right)^{-1}.
\label{eqn:MStepH}
\end{align}
%
In the  linear Gaussian setting where the involved matrices are non-singular, this procedure monotonically increases the likelihood (for $\epsilon=0$) until convergence \cite{Dempster1977}, but typically requires multiple forward–backward passes (violating~\ref{itm:latency}).

{\bf RTSNet:} 
The {\em RTSNet} algorithm proposed in \cite{ni2022rtsnet} is a neural-augmented   \ac{rts} smoother based on the KalmanNet methodology~\cite{revach2022kalmannet}. It replaces the computation of the forward and backward gains, $\Kgain_t$ and $\Sgain_t$ in \eqref{eq:FW_update_1}-\eqref{eq:BW_update_1}, with dedicated \acp{rnn} trained jointly in an end-to-end manner, without explicitly tracking the second-order moments.
The forward pass follows the  \ac{ekf} and the backward pass applies the update of \eqref{eq:BW_update_1}, while the learned gain computations enable coping with approximated \ac{ss} models (Challenge \ref{itm:Approx}) and non-Gaussian dynamics (Challenge \ref{itm:Noise}). However, as the learned gains are tied to a fixed training distribution, RTSNet cannot readily accommodate time-varying \ac{ss} models (Challenge \ref{itm:varying}).


\section{EM-KalmanNet}
\label{sec:method}

\begin{figure}
	\centering
	\includegraphics[width=0.9\linewidth]{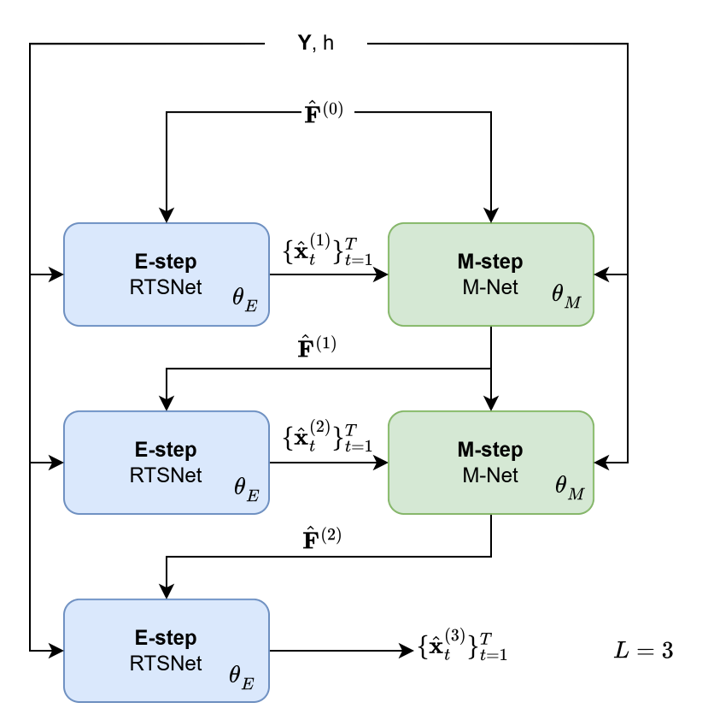}
    \caption{EM-KalmanNet with $\boldsymbol{\phi}=\myMat{F}$ and $L=3$ iterations.}

	\label{fig:EM}
\end{figure}

This section introduces the EM-KalmanNet algorithm, illustrated in Fig.~\ref{fig:EM}. 
The design of EM-KalmanNet is motivated by combining the complementary strengths of the model-based EM-KF and of AI-aided KalmanNet-type smoothers to jointly address Challenges \ref{itm:Approx}–\ref{itm:shortBlocks}. The EM-KF framework provides a principled mechanism for estimating unknown model parameters alongside the latent states, making it suited to address the unknown block-wise variations in~\ref{itm:varying}. However, it relies on linear-Gaussian assumptions, typically requires many iterations until convergence, and benefits from long observation blocks, rendering it unsuitable for Challenges \ref{itm:Approx}–\ref{itm:Noise},  \ref{itm:latency}, and \ref{itm:shortBlocks}.  RTSNet inherits from KalmanNet the ability to operate under mismatched models and non-Gaussian noise (Challenges~\ref{itm:Approx}–\ref{itm:Noise}), but cannot readily adapt to varying dynamics at inference without labeled online data.

Rather than learning an entirely new tracking architecture, EM-KalmanNet unfolds the EM iterations into a trainable architecture, replacing the analytical E- and M-step computations with learnable modules while preserving the EM-inspired estimation flow. 
The learned E-step is implemented by the parameter-aware RTSNet,
with trainable parameters $\theta_E$, while the learned M-step is
implemented by M-Net, with trainable parameters $\theta_M$.
The separation into two modules preserves the interpretable structure of the EM-KF, allows each module to specialize in its respective task, and enables the fixed, small iteration budget to satisfy the latency requirement in Challenge~\ref{itm:latency}. We next detail the architecture in Subsection~\ref{subsec:Algorithm} and its training in Subsection~\ref{subsec:training}, and conclude with a discussion in Subsection~\ref{subsec:Discussion}.

\subsection{EM-KalmanNet Algorithm}
\label{subsec:Algorithm}
We next detail the construction of EM-KalmanNet. For 
simplicity, we omit the block index $i$ and describe the processing of a single block of $T$ observations.

\smallskip
{\bf Initialization:}
EM-KalmanNet operates with an initial estimate $\hat{\boldsymbol{\phi}}^{(0)}$ of the unknown model parameter. When processing consecutive blocks, $\hat{\boldsymbol{\phi}}^{(0)}$ is set to the parameter estimate obtained for the previous block, exploiting the gradual block-wise variation of the model.  For the first block, $\hat{\boldsymbol{\phi}}^{(0)}$ can be obtained from prior model knowledge or from an offline estimate computed from the training data.

\smallskip
{\bf Parameter-Aware RTSNet (Learned E-Step):}
A key requirement of EM-KalmanNet is that the smoother used in the E-step operates reliably under different values of the unknown parameter, since the parameter estimate is refined across the unfolded iterations. Standard RTSNet is trained for a fixed \ac{ss} model, and its learned gains implicitly encode that specific model; it therefore cannot directly account for variations in $\myMat{F}$ or $\myMat{H}$. We thus employ a \emph{parameter-aware} RTSNet, denoted ${\rm RTSNet}_{\rm PA}$, in which the current parameter estimate $\hat{\boldsymbol{\phi}}^{(l-1)}$ is provided as an additional feature to the \acp{rnn} that compute the forward and backward gains. 

\smallskip
{\bf M-Net (Learned M-Step):}
The M-step of EM-KalmanNet is implemented by a neural module termed {\em M-Net}, which learns to update the unknown model parameter from statistics computed using the smoothed trajectory. Rather than directly outputting the next parameter estimate, M-Net predicts a correction to the current estimate,
\begin{align}
    \Delta\boldsymbol{\phi}^{(l)}
    &=
    {\rm MNet}(\myVec{z}_{\boldsymbol{\phi}}^{(l)};\myVec{\theta}_{\rm M}), \quad
    \hat{\boldsymbol{\phi}}^{(l)}
    =
    \hat{\boldsymbol{\phi}}^{(l-1)}
    +
    \Delta\boldsymbol{\phi}^{(l)},
    \label{eqn:MNetUpdate}
\end{align}
where $\myVec{z}_{\boldsymbol{\phi}}^{(l)}$ is a feature vector (whose structure is detailed below), and $\myVec{\theta}_{\rm M}$ denotes the trainable parameters of M-Net. In our numerical study in Section~\ref{sec:numerical_study}, we implemented M-Net as a fully-connected \ac{dnn} with GELU activations and a scaled hyperbolic tangent output layer to produce a bounded correction $\Delta\boldsymbol{\phi}^{(l)}$. 
The correction-based update  \eqref{eqn:MNetUpdate} is motivated by the block-wise setting, where the model varies gradually across consecutive blocks.  A bounded local correction around the current estimate is easier to learn reliably than a full re-estimation, particularly under short block lengths (\ref{itm:shortBlocks}).

\smallskip
{\bf Features Design:}
The input $\myVec{z}_{\boldsymbol{\phi}}^{(l)}$  comprises two types of quantities: $(i)$ empirical counterparts of the EM-KF sufficient statistics \eqref{eqn:A_def}, which describe the correlation structure needed for parameter estimation; and $(ii)$ residual- and gradient-based statistics, computed after applying the current parameter estimate, which indicate how this estimate should be corrected.
Since RTSNet does not explicitly track the posterior covariance matrices
$\hat{\myMat{P}}_{t}^{(l)}$ and $\hat{\myMat{P}}_{t,t-1}^{(l)}$, the covariance-dependent statistics of \eqref{eqn:A_def} are replaced with empirical moments of the smoothed trajectory.
This statistics-based design avoids feeding the entire trajectory into the M-step network, yielding a compact, block-length-independent update rule that preserves the structure of the classical EM. 
 
We next formulate how $\myVec{z}_{\boldsymbol{\phi}}^{(l)}$  is designed for each of the considered cases. For brevity, we define the $T$-length empirical cross-covariance of time sequences $\{\myVec{a}_t\}$ and  $\{\myVec{b}_t\}$  as 
\begin{equation*}
    {\rm Cor}_T\{\myVec{a}_t, \myVec{b}_t\} \triangleq\frac{1}{T}\sum_{t=1}^T \Big(\myVec{a}_t -\frac{1}{T}\sum_{\tau=1}^T \myVec{a}_\tau\Big)  \Big(\myVec{b}_t - \frac{1}{T}\sum_{\tau=1}^T \myVec{b}_\tau\Big)^{\top}. 
\end{equation*}

\emph{Case 1 (Unknown-$\myMat{F}$):}
 For this case, we define the one-step prediction under the current estimate and its residual as
\begin{align}
\label{eqn:transResidual}
\myVec{x}_{t}^{-,(l)}
\triangleq
\hat{\myMat{F}}^{(l-1)}\hat{\myVec{x}}_{t-1}^{(l)}, ~~~
\Delta\myVec{x}_{t}^{(l)}
\triangleq
\hat{\myVec{x}}_{t}^{(l)}-\myVec{x}_{t}^{-,(l)} ,
\end{align}
and the observation residual under the available observation function as $
\myVec{\nu}_{F,t}^{(l)}
\triangleq
\myVec{y}_t-h_t(\hat{\myVec{x}}_{t}^{(l)})$. 
We form the transition- and observation-residual covariances 
$\myMat{S}_{\Delta x}^{(l)}
\triangleq{\rm Cor}_T\{\Delta\myVec{x}_{t}^{(l)}, \Delta\myVec{x}_{t}^{(l)}\}$ and
$\myMat{S}_{\nu,F}^{(l)} 
\triangleq{\rm Cor}_T\{\myVec{\nu}_{F,t}^{(l)}, \myVec{\nu}_{F,t}^{(l)}\}$. 
The matrix $\myMat{S}_{\Delta x}^{(l)}$ measures the empirical covariance of the transition errors induced by the current estimate $\hat{\myMat{F}}^{(l-1)}$: when this estimate is accurate, these errors mainly reflect the process noise, whereas an inaccurate transition matrix inflates them with a structured mismatch component.  Similarly, $\myMat{S}_{\nu,F}^{(l)}$ quantifies the remaining error in the measurement domain.

Next, we use the cross-statistic  
$\myMat{C}_{\Delta x,x}^{(l)}
\triangleq
\frac{1}{T}\sum_{t=1}^{T}
\Delta\myVec{x}_{t}^{(l)}
\left(\hat{\myVec{x}}_{t-1}^{(l)}\right)^\top$,
 which provides directional information for correcting $\myMat{F}$, as 
 \begin{equation*}
     \myMat{C}_{\Delta x,x}^{(l)} = -\left.
\nabla_{\myMat{F}} \left(\frac{1}{2T}\sum_{t=1}^{T}
\left\|
\hat{\myVec{x}}_{t}^{(l)}
-
\myMat{F}\hat{\myVec{x}}_{t-1}^{(l)}
\right\|^2\right) 
\right|_{\myMat{F}=\hat{\myMat{F}}^{(l-1)}}.
 \end{equation*}
 The resulting M-Net input for the unknown-$\myMat{F}$ case is
\begin{align}
\label{eqn:zF}
\myVec{z}_{\myMat{F}}^{(l)}
=
\Big[
&({\rm vec}(\tilde{\myMat{A}}_1^{(l)}))^\top,
({\rm vec}(\tilde{\myMat{A}}_2^{(l)}))^\top,
({\rm vec}(\myMat{S}_{\Delta x}^{(l)}))^\top,
({\rm vec}(\myMat{S}_{\nu,F}^{(l)}))^\top, \notag\\
&({\rm vec}(\myMat{C}_{\Delta x,x}^{(l)}))^\top,
({\rm vec}(\hat{\myMat{F}}^{(l-1)}))^\top
\Big]^\top,
\end{align}
where 
\begin{align}
\tilde{\myMat{A}}_1^{(l)}
&\triangleq
\sum_{t=1}^{T}
\hat{\myVec{x}}_{t}^{(l)}
(\hat{\myVec{x}}_{t-1}^{(l)})^\top, 
~~~
\tilde{\myMat{A}}_2^{(l)}
&\triangleq
\sum_{t=1}^{T}
\hat{\myVec{x}}_{t-1}^{(l)}
(\hat{\myVec{x}}_{t-1}^{(l)})^\top .
\end{align}

\emph{Case 2 (Unknown-$\myMat{H}$):}
Analogously, for this case, the observation residual under the current estimate is
\begin{align}
\myVec{\nu}_{H,t}^{(l)}
&\triangleq
\myVec{y}_t-\hat{\myMat{H}}^{(l-1)}\hat{\myVec{x}}_{t}^{(l)} 
\end{align}
from which we form the residual covariance $\myMat{S}_{\nu,H}^{(l)} 
\triangleq{\rm Cor}_T\{\myVec{\nu}_{H,t}^{(l)}, \myVec{\nu}_{H,t}^{(l)}\}$ and residual-state cross-statistic $
\myMat{C}_{\nu x}^{(l)}
\triangleq
\frac{1}{T}\sum_{t=1}^{T}
\myVec{\nu}_{H,t}^{(l)}
(\hat{\myVec{x}}_{t}^{(l)})^\top$. 
As in Case 1, $\myMat{S}_{\nu,H}^{(l)}$ captures the covariance structure of the observation residuals induced by the current estimate $\hat{\myMat{H}}^{(l-1)}$, and therefore provides an empirical description of the remaining measurement-domain error. 
The cross-statistic $\myMat{C}_{\nu x}^{(l)}$  is the negative gradient of the observation fitting loss $\frac{1}{2T}\sum_{t=1}^{T}
\left\|
\myVec{y}_{t}
-
\myMat{H}\hat{\myVec{x}}_{t}^{(l)}
\right\|^2$  
evaluated at $\hat{\myMat{H}}^{(l-1)}$, and thus encodes the steepest-descent
correction direction.

 The M-Net input for the unknown-$\myMat{H}$ case is
\begin{align}
\label{eqn:zH}
\myVec{z}_{\myMat{H}}^{(l)}
=
\Big[
&({\rm vec}(\myMat{A}_4^{(l)}))^\top,
({\rm vec}(\tilde{\myMat{A}}_3^{(l)}))^\top,
({\rm vec}(\myMat{S}_{\nu,H}^{(l)}))^\top, \notag\\
&~({\rm vec}(\myMat{C}_{\nu x}^{(l)}))^\top,
({\rm vec}(\hat{\myMat{H}}^{(l-1)}))^\top
\Big]^\top,
\end{align}
where 
\begin{align}
\tilde{\myMat{A}}_3^{(l)}
&\triangleq
\sum_{t=1}^{T}
\hat{\myVec{x}}_{t}^{(l)}
(\hat{\myVec{x}}_{t}^{(l)})^\top, 
\end{align}

\begin{remark}[Relation to the analytical EM update]
\label{rem:EMspecial}
The M-Net features in \eqref{eqn:zF} and \eqref{eqn:zH} are chosen such that the covariance-agnostic counterpart of the classical M-step can be expressed as a correction of the form in \eqref{eqn:MNetUpdate}.
Consider the unknown-$\myMat{H}$ case. By the definition of $\myVec{\nu}_{H,t}^{(l)}$,
\begin{align}
\myMat{A}_4^{(l)}
=
T\myMat{C}_{\nu x}^{(l)}
+
\hat{\myMat{H}}^{(l-1)}
\tilde{\myMat{A}}_3^{(l)}.
\end{align}
Substituting this into the update \eqref{eqn:MStepH}, with $\myMat{A}_3^{(l)}$ replaced by $\tilde{\myMat{A}}_3^{(l)}$, yields
\begin{equation}
\hat{\myMat{H}}^{(l)}
=
\hat{\myMat{H}}^{(l-1)}
+
\big(
T\myMat{C}_{\nu x}^{(l)}
-
\epsilon
\hat{\myMat{H}}^{(l-1)}
\big)
\big(\tilde{\myMat{A}}_3^{(l)}+
\epsilon\myMat{I}_m
\big)^{-1}.
\end{equation}
Thus, the update is a preconditioned gradient-type correction whose ingredients, $\myMat{C}_{\nu x}^{(l)}$, $\tilde{\myMat{A}}_3^{(l)}$, and $\hat{\myMat{H}}^{(l-1)}$, are all contained in $\myVec{z}_{\myMat{H}}^{(l)}$.
 An analogous identity holds for the unknown-$\myMat{F}$ case, using $\tilde{\myMat{A}}_1^{(l)}=T\myMat{C}_{\Delta x,x}^{(l)}+\hat{\myMat{F}}^{(l-1)}\tilde{\myMat{A}}_2^{(l)}$ together with  \eqref{eqn:MStepF}.
 The only approximation involved is the replacement of the exact EM statistics with their empirical counterparts, which omits the posterior covariance terms $\hat{\myMat{P}}_t^{(l)}$, $\hat{\myMat{P}}_{t,t-1}^{(l)}$ that RTSNet does not explicitly track. Consequently, M-Net can in principle reproduce this covariance-agnostic analytical EM update as a special case of \eqref{eqn:MNetUpdate}, while its learned nonlinear mapping is free to realize more general corrections, adapted to non-Gaussian noise and model mismatch.
\end{remark} 


\smallskip
{\bf Overall Algorithm:}
EM-KalmanNet is obtained by unfolding the EM-KF flow into $L$ EM-like iterations, where $L$ is a small integer set to satisfy the latency requirement of Challenge~\ref{itm:latency}
(e.g., $L=3$ as used in Section~\ref{sec:numerical_study}).
 In the first $L-1$ iterations, the E-step is implemented by the shared RTSNet and is followed by an M-step implemented by the shared M-Net. In the final iteration, only the E-step is performed, so that EM-KalmanNet uses $L$ RTSNet passes and $L-1$ M-Net updates. The per-block processing is summarized in Algorithm~\ref{alg:EmKalmanNet}.
When operating over multiple blocks, as formulated in Subsection~\ref{ssec:SS}, the initialization $\hat{\boldsymbol{\phi}}^{(0)}$ of the $i$th block is set to the estimate $\hat{\boldsymbol{\phi}}^{(L-1)}$  produced for block $i-1$. 
The initial state estimate is propagated from the previous block, whereas its covariance is reinitialized at the beginning of each block.

\vspace{-0.25cm}
\begin{algorithm}
\caption{EM-KalmanNet at one block}
\label{alg:EmKalmanNet} 
\SetKwInOut{Initialization}{Init}
\Initialization{Trained RTSNet ($\myVec{\theta}_{\rm E}$); trained M-Net ($\myVec{\theta}_{\rm M}$)} 
\SetKwInOut{Input}{Input}
\Input{Observation $\{\myVec{y}_t\}_{t=1}^T$; initial estimate $\hat{\boldsymbol{\phi}}^{(0)}$; initial state moments; available model relation $h(\cdot)$ (unknown-$\mathbf{F}$ case) or $f(\cdot)$ (unknown-$\mathbf{H}$ case);}
\For{$l=1,\ldots, L-1$}
{
 \nonl\textbf{E-Step:}\\
 Set $\{\hat{\myVec{x}}_{t}^{(l)}\}_{t=1}^{T} \gets
{\rm RTSNet}_{\rm PA}(\{\myVec{y}_t\}_{t=1}^{T}, \hat{\boldsymbol{\phi}}^{(l-1)}; \myVec{\theta}_{\rm E})$\;

  \nonl\textbf{M-Step:}\\
Set $\myVec{z}^{(l)}_{\phi}$ via \eqref{eqn:zF} or \eqref{eqn:zH} ;\\
 Obtain correction $\Delta\boldsymbol{\phi}^{(l)}
\leftarrow \mathrm{MNet}(\myVec{z}^{(l)}_{\phi};\boldsymbol{\theta}_M)$;\\
Update parameters $\hat{\boldsymbol{\phi}}^{(l)}
\leftarrow \hat{\boldsymbol{\phi}}^{(l-1)}+\Delta\boldsymbol{\phi}^{(l)}$;
}
\nonl\textbf{Final E-Step:}\\
Set $\{\hat{\myVec{x}}_{t}^{(L)}\}_{t=1}^{T} \gets {\rm RTSNet}_{\rm PA}(\{\myVec{y}_t\}_{t=1}^{T}, \hat{\boldsymbol{\phi}}^{(L-1)}; \myVec{\theta}_{\rm E})$\;
\KwRet{$\{\hat{\myVec{x}}_{t}^{(L)}\}_{t=1}^{T}$, $\hat{\boldsymbol{\phi}}^{(L-1)}$}
\end{algorithm}

\vspace{-0.75cm}
\subsection{Training}
\label{subsec:training}
\vspace{-0.1cm}

Algorithm~\ref{alg:EmKalmanNet} requires setting the parameters of the shared RTSNet, $\myVec{\theta}_{\rm E}$, and of the shared M-Net, $\myVec{\theta}_{\rm M}$.  These parameters are learned offline from the dataset $\mySet{D}$ in \eqref{eq:dataset} via a three-stage training procedure, summarized in Algorithm~\ref{alg:EmKalmanNetTraining}. The staged design first establishes a reliable smoother, then a reliable parameter-update rule on top of it, and finally adapts both modules to operate jointly through the unfolded iterations; this ordering avoids the optimization instabilities that arise when both modules are trained jointly from scratch.

\smallskip 
{\bf Stage 1: Pre-Train RTSNet.}
We first train the parameter-aware RTSNet to recover the latent state trajectory from the observations given a model-parameter value provided as input. For each training block, indexed by $(j,i)$, let ${\boldsymbol{\phi}}_{j,i}$ denote the parameter value provided to
RTSNet for conditioning. Given the observations $\{\mathbf{y}_{j,i,t}\}_{t=1}^{T}$ and
${\boldsymbol{\phi}}_{j,i}$, RTSNet is trained with the state-estimation loss
\begin{align}
\mathcal{L}_{\mySet{D}}^{\rm RTSNet}(\myVec{\theta}_{\rm E})
=
\frac{1}{N_{\mathcal D}MT}
\sum_{j=1}^{N_{\mathcal D}}
\sum_{i=1}^{M}
\sum_{t=1}^{T}
\big\|
\hat{\myVec{x}}_{j,i,t}-
\myVec{x}_{j,i,t}
\big\|_2^2 ,
\end{align}
where 
$\hat{\myVec{x}}_{j,i,t}$ is the smoothed state produced by 
${\rm RTSNet}_{\rm PA}(\{\myVec{y}_{j,i,\tau}\}_{\tau=1}^{T},\boldsymbol{\phi}_{j,i};\myVec{\theta}_{\rm E})$, and $\myVec{\theta}_{\rm E}$ are the current network parameters.

 To make the smoother robust to the imperfect parameter estimates encountered during the unfolded  iterations, Stage~1 comprises two phases that differ in the  conditioning parameter $\boldsymbol{\phi}_{j,i}$: in the first phase we set $\boldsymbol{\phi}_{j,i}=\boldsymbol{\phi}_{j,i}^{\star}$, i.e., the ground-truth parameter, so that RTSNet learns accurate smoothing under the correct model;  in the second phase we set $\boldsymbol{\phi}_{j,i}=\boldsymbol{\phi}_{j,i}^{\star}+\delta\boldsymbol{\phi}_{j,i}$,
where $\delta\boldsymbol{\phi}_{j,i}$ is a random perturbation whose magnitude is matched to the errors anticipated at inference.  
The second phase deliberately exposes RTSNet to mismatched inputs, teaching it to remain informative when conditioned on inexact parameter estimates.

\smallskip
 {\bf Stage 2: Train M-Net with Frozen RTSNet.}
Given the pre-trained RTSNet, we train the M-Net to update the unknown parameter while keeping the RTSNet parameters $\myVec{\theta}_{\rm E}$ fixed. At each unfolded iteration, RTSNet produces a smoothed trajectory, from which the feature vector $\myVec{z}_{\boldsymbol{\phi}}^{(l)}$ 
is computed as described above. The M-Net then predicts an additive correction $\Delta\boldsymbol{\phi}^{(l)}$. 
The Stage-2 loss combines a parameter-estimation error, a state-estimation term, and an update regularizer
%
\begin{eqnarray}
\mathcal{L}_{\mySet{D}}^{\rm M}(\myVec{\theta}_{\rm M})
\hspace{6.5cm}\nonumber\\
=
\frac{1}{N_{\mathcal D}M}
\sum_{j=1}^{N_{\mathcal D}}
\sum_{i=1}^{M}
\Bigg[
\sum_{l=1}^{L}
a_l
\lambda_x^{\rm M}\cdot
\frac{1}{T}
\sum_{t=1}^{T}
\left\|
\hat{\myVec{x}}_{j,i,t}^{(l)}
-
\myVec{x}_{j,i,t}
\right\|_2^2\nonumber
\\
\quad+
\sum_{l=1}^{L-1}
a_l
\left(
\lambda_{\phi}^{\rm M}
\left\|
\hat{\boldsymbol{\phi}}_{j,i}^{(l)}
-
\boldsymbol{\phi}_{j,i}^{\star}
\right\|_F^2
+
\lambda_{\Delta}^{\rm M}
\left\|
\Delta\boldsymbol{\phi}_{j,i}^{(l)}
\right\|_F^2
\right)
\Bigg],
\label{eqn:mLoss}
\end{eqnarray}
with hyperparameters $\lambda_{\phi}^{\rm M}\gg\lambda_x^{\rm M},\lambda_{\Delta}^{\rm M}\geq 0$. Since the goal of this stage is to learn a reliable M-step update, the parameter-fit weight  $\lambda_{\phi}^{\rm M}$ is chosen to be dominant.  The regularizer $\lambda_{\Delta}^{\rm M}\|\Delta\boldsymbol{\phi}\|_{\rm F}^2$ discourages excessively large corrections, complementing the bounded output activation of M-Net.
The per-iteration weights satisfy $a_l>0$ and $\sum_{l=1}^{L}a_l=1$, and are chosen to be increasing in $l$ so as to emphasize the later, more refined iterations while still providing a training signal to the early ones.   

\smallskip
 {\bf Stage 3: Joint Unfolded Fine-Tuning.}
In the final stage, RTSNet and M-Net are trained jointly through the unfolded iterations, adapting the smoother and the learned M-step to operate in concert. The loss retains the structure of \eqref{eqn:mLoss} from Stage 2, where 
$\lambda_x^{\rm M}, \lambda_{\phi}^{\rm M}$, and $\lambda_{\Delta}^{\rm M}$ are replaced by
$\lambda_x^{\rm E2E}$, $
\lambda_{\phi}^{\rm E2E}$, and $
\lambda_{\Delta}^{\rm E2E}$, respectively, where now $\lambda_x^{\rm E2E}$ is chosen to be dominant, since the ultimate objective of EM-KalmanNet is accurate state smoothing rather than parameter recovery. The parameter-fit term acts as an auxiliary signal guiding the adaptation.

To encourage each module to learn a robust update rule from the statistics it receives, rather than relying on the gradient flow across iterations, the gradients are detached between consecutive EM-KalmanNet iterations during training. Thus, at each iteration $l$, RTSNet and M-Net are trained to optimally process their inputs given the output of iteration $l-1$ treated as fixed, effectively training each iteration as a self-contained estimation step.  Beyond stabilizing training, this design has an important practical benefit: since each iteration is trained to improve upon an arbitrary incoming estimate, the trained modules remain valid when the number of iterations applied at inference deviates from $L$, allowing the iteration budget to be traded off against latency at deployment.

\begin{algorithm}
\caption{EM-KalmanNet Training}
\label{alg:EmKalmanNetTraining} 
\SetKwInOut{Input}{Input}
\Input{Dataset $\mySet{D}$; number of unfolded iterations $L$}
\SetKwInOut{Output}{Output}
\Output{Trained parameters $\myVec{\theta}_{\rm E}$ and  $\myVec{\theta}_{\rm M}$}

\nonl\textbf{Stage 1: Pre-Train RTSNet}\\
Train $\myVec{\theta}_{\rm E}$ on $\mySet{D}$ using 
$\mathcal{L}_{\mySet{D}}^{\rm RTSNet}$, 
first with $\boldsymbol{\phi}_{j,i}=\boldsymbol{\phi}_{j,i}^{\star}$,
then with perturbed parameters
$\boldsymbol{\phi}_{j,i}=\boldsymbol{\phi}_{j,i}^{\star}+\delta\boldsymbol{\phi}_{j,i}$\;
\nonl\textbf{Stage 2: Train M-Net with Frozen RTSNet}\\
Freeze $\myVec{\theta}_{\rm E}$\;
\For{each mini-batch}{
  Run Algorithm~\ref{alg:EmKalmanNet} with mismatched initialization $\boldsymbol{\phi}^{(0)}$\;

Update $\myVec{\theta}_{\rm M}$ using \eqref{eqn:mLoss}\;}

\nonl\textbf{Stage 3: Joint Unfolded Fine-Tuning}\\
Unfreeze $\myVec{\theta}_{\rm E}$\;
\For{each mini-batch}{
  \hspace{-0.22cm}Run Algorithm~\ref{alg:EmKalmanNet}, detaching gradients between iterations\;

Jointly update $\myVec{\theta}_{\rm E}$ and $\myVec{\theta}_{\rm M}$ 
using $\mathcal{L}_{\mySet{D}}^{\rm E2E}$\;}

\KwRet{$\myVec{\theta}_{\rm E}, \myVec{\theta}_{\rm M}$}
\end{algorithm}

%

\subsection{Discussion}
\label{subsec:Discussion}

{\bf Tackling the Formulated Problem:}
EM-KalmanNet addresses Challenges \ref{itm:Approx}–\ref{itm:shortBlocks} 
through dedicated components: the RTSNet-based E-step handles mismatched dynamics (\ref{itm:Approx}) and unknown, non-Gaussian noise (\ref{itm:Noise}); the parameter-aware modulation combined with the residual M-Net update tracks block-wise variations 
(\ref{itm:varying}); the unfolded structure with a small fixed $L$ ensures bounded latency (\ref{itm:latency}); and the bounded, correction-based parameter update, together with the warm-starting of $\boldsymbol{\phi}^{(0)}$ across blocks, enables reliable adaptation from short blocks (\ref{itm:shortBlocks}), as a local refinement is learnable from far fewer samples than a full re-estimation.

\smallskip
\textbf{Relation to Classical EM-KF:}
Several special cases clarify the connection between EM-KalmanNet and classical EM-KF.
If RTSNet is replaced by the exact RTS smoother and M-Net by the analytical EM M-step in \eqref{eqn:MStepF} or \eqref{eqn:MStepH}, the unfolded architecture reduces to EM-KF truncated after $L$ iterations.
 By Remark~\ref{rem:EMspecial}, this analytical M-step is itself approximately realizable within the learned architecture, up to the omitted posterior-covariance terms. 
Conversely, setting $\Delta\boldsymbol{\phi}^{(l)}=\mathbf{0}$ reduces EM-KalmanNet to RTSNet operating with a fixed model parameter. EM-KalmanNet can thus be viewed as an EM-inspired unfolded architecture that bridges model-based EM-KF and supervised end-to-end tracking, trading the asymptotic convergence guarantees of EM (which hold only under linear-Gaussian assumptions) for learned, fixed-complexity adaptation applicable to the general model \eqref{eqn:ssModel}.

\smallskip
{\bf Computational Complexity:}
For a block of $T$ measurements, EM-KalmanNet performs $L$
forward--backward passes of RTSNet and $L-1$ parameter updates by
M-Net. Each RTSNet pass requires $\mathcal{O}(T c_{\rm E})$
operations, where $c_{\rm E}$ denotes the per-time-step computational
cost of RTSNet, including the learned gain networks and the available
model functions. Each M-Net update requires
$\mathcal{O}(T c_{\rm S}+c_{\rm M})$ operations, where $c_{\rm S}$
denotes the per-time-step cost of accumulating the required summary
statistics and $c_{\rm M}$ is the cost of evaluating M-Net.
Consequently, the overall per-block complexity is of order $
\mathcal{O}\!\left(
LTc_{\rm E}
+ (L-1)(Tc_{\rm S}+c_{\rm M})
\right)$.
For fixed network architectures and state and observation dimensions,
the complexity scales linearly with both the block length $T$ and the
unfolded depth $L$. Since $L$ is fixed at design time, the
computational budget is known in advance, facilitating predictable
low-latency inference as required by Challenge~\ref{itm:latency}.

\smallskip
{\bf Future Research:}
Several directions remain open for future study. First, the current framework can be extended to simultaneous adaptation of multiple model components, such as the dynamics and observation model, as well as to time-varying noise statistics. Second, online or streaming variants may be developed by applying the learned update mechanism over sliding windows rather than fixed blocks. Third, integration with hypernetworks~\cite{ni2024adaptive} may allow the adaptation modules themselves to change according to the operating regime, further improving generalization across different forms of time variation and model mismatch. These aspects are left for future study.


\section{Experimental Study}
\label{sec:numerical_study}
We next evaluate EM-KalmanNet in several \ac{ss} tracking scenarios\footnote{Our source code and all hyperparameters can be found
at \url{https://github.com/oricoohen/EMKF_NET.git}}.  We first describe the common experimental setup  (Subsection~\ref{subsec:exp_setup}). Then, we evaluate EM-KalmanNet on controlled synthetic \ac{ss} models with an unknown time-varying state-transition matrix $\myMat{F}$  (Subsection~\ref{subsec:synthetic_F}) and with an unknown time-varying observation matrix $\myMat{H}$ (Subsection~\ref{subsec:synthetic_H}), followed by a nonlinear Lorenz attractor model (Subsection~\ref{subsec:lorenz}) and an acoustic source-tracking application (Subsection~\ref{subsec:acoustic}). 

 Each experiment isolates one or more of Challenges~\ref{itm:Approx}--\ref{itm:shortBlocks}: nonlinear and chaotic dynamics test robustness to model approximation (\ref{itm:Approx}); non-Gaussian and correlated observation noise test robustness to distributional mismatch (\ref{itm:Noise}); block-wise varying $\myMat{F}$ or $\myMat{H}$ test adaptation to unknown time variation (\ref{itm:varying}); reported inference latencies test computational efficiency (\ref{itm:latency}); and a dedicated sweep over the block length $T$ tests operation under rapidly varying, sparsely observed conditions (\ref{itm:shortBlocks}).

\subsection{Experimental Setup}
\label{subsec:exp_setup}

{\bf Compared Algorithms:}
We compare EM-KalmanNet with several model-based and data-driven benchmarks.
The model-based benchmarks include the \ac{rts} smoother \cite{rauch1965maximum} with full knowledge of the \ac{ss} model at each block ({\em RTS Full}), the \ac{rts} smoother that uses a single fixed, mismatched parameter for all blocks ({\em RTS False}), and the model-based EM-KF\cite{Shumway1982}, initialized with the same mismatched parameter as EM-KalmanNet.
As data-driven benchmarks, we evaluate RTSNet\cite{ni2022rtsnet} operating with the same fixed mismatched parameter ({\em RTSNet False}), and a bidirectional GRU (BGRU)\cite{cho2014learning} smoother as a purely data-driven sequence-to-sequence benchmark.
In the synthetic state-transition adaptation experiment (Subsection~\ref{subsec:synthetic_F}), we also compare with our preliminary M-step implementation ({\em M-RTSNETP}) \cite{Cohen2026EMKalmanNet} to isolate the gains of the proposed architecture and training procedure.
 
Within each experiment, all data-driven methods are trained using the same training dataset of $N_{\mathcal D}=400$ trajectories, each consisting of $M$ blocks of length $T$, where $M$ and $T$ are experiment-specific and stated in each subsection.
EM-KalmanNet is trained using the unfolded procedure described in Subsection~\ref{subsec:training}, with the parameter-aware RTSNet and M-Net being shared across iterations. For a controlled comparison, both adaptive methods use a small, fixed iterative budget of $3$ by default. This fixed computational budget is consistent with the latency requirement of Challenge~\ref{itm:latency}; Subsection~\ref{subsec:lorenz} additionally reports an EM-KF variant with a larger budget for reference.


\smallskip
{\bf Performance Metrics and Noise Levels:}
Performance is evaluated using the state-estimation \ac{mse} in dB,
averaged over the test trajectories and time indices.
In all experiments except the observation-noise-distribution
experiment, the process- and observation-noise covariance matrices
are parameterized as
$\myMat{Q}=q^2\myMat{I}_m$ and
$\myMat{R}=r^2\myMat{I}_n$, respectively, where $q^2$ and $r^2$
control the corresponding noise levels. Depending on the experiment,
either $q^2$ is held fixed while $r^2$ is varied, or the ratio
$q^2/r^2$ is kept fixed while both are scaled jointly.
Accordingly, when results are reported versus $1/r^2$ in dB, larger
values correspond to cleaner observations.

\subsection{Synthetic State-Transition Adaptation (Case 1)}
\label{subsec:synthetic_F}
\subsubsection{Setup}
In all three experiments, each test trajectory consists of three blocks of length $T=30$. The reference dynamics is set by the nominal state-transition matrix, 
\begin{align}
\myMat{F}_0 =
\begin{bmatrix}
0.83 & 0.20\\
0.20 & 0.83
\end{bmatrix}.
\end{align}
The state-transition matrix of the first block is obtained by rotating $\myMat{F}_0$ by $0.2$ radians. Each subsequent block applies an additional $0.2$-radian rotation to that of the preceding block.
In this experiment, we fix $q^2=0.01$ and vary $r^2$. Thus, the
reported $1/r^2$ axis in dB directly reflects changes in the
observation \ac{snr}.
The three experiments differ only in the observation model: $(i)$ a linear
observation model
\begin{align}
\myMat{H}=
\begin{bmatrix}
1 & 1\\
0.25 & 1
\end{bmatrix}
\label{eqn:Hmat}
\end{align}
with Gaussian noise; $(ii)$ the same  $\myMat{H}$ with exponentially distributed noise; and $(iii)$   a nonlinear observation model
\begin{align}
h(\myVec{x})
=
\myMat{H}\myVec{x}
+
0.3h_0(\myVec{x}), \quad
h_0(\myVec{x})
=
\begin{bmatrix}
\sqrt{x_1^2+x_2^2}\\
\operatorname{atan2}(x_2,x_1)
\end{bmatrix}.
\end{align}

The training trajectories follow the same three-block structure as the test trajectories. The first block uses $\myMat{F}_0$ rotated by angle sampled from $\mathcal{U}[-1,1]$ radians. Each subsequent state-transition matrix is obtained by rotating using a newly
sampled rotation angle from the same distribution. This exposes the data-driven methods to a range of block-wise model variations during training.
 

\subsubsection{Linear Gaussian Setting}
\begin{figure}
\centering
\begin{tikzpicture}
\begin{axis}[
    width=\linewidth,
    height=0.80\linewidth,
    xmin=-10, xmax=30,
    ymin=-30, ymax=7,
    xtick={-10,0,10,20,30},
    ytick={-30,-20,-10,0,10},
    grid=both,
    xlabel={$\frac{1}{r^{2}}\,[\mathrm{dB}]$},
    ylabel={MSE [dB]},
    label style={font=\small},
    tick label style={font=\small},
    every axis plot/.append style={line width=1.0pt, mark size=2.4pt},
    legend columns=2,
    clip=false,
    legend style={
        at={(0.02,0.02)},
        anchor=south west,
        draw=none,
        fill=white,
        fill opacity=0.80,
        text opacity=1,
        font=\tiny,
        rounded corners=1pt
    },
    legend cell align=left
]

\addplot+[black, mark=triangle*]
  coordinates {(-10,-11.2) (0,-16) (10,-18.8) (20,-22) (30,-27.6)};
\addlegendentry{RTS Full};

\addplot+[orange, dashed, mark=triangle*]
  coordinates {(-10,4.8) (0,4.2) (10,-0.54) (20,-14.7) (30,-27.5)};
\addlegendentry{EM-KF};

\addplot+[teal!70!black, mark=o]
  coordinates {(-10,4.9) (0,4.7) (10,3.7) (20,-2.48) (30,-17.39)};
\addlegendentry{RTS False};

\addplot+[violet, mark=o]
  coordinates {(-10,5.4) (0,1.1) (10,-5.1) (20,-13.5) (30,-19.5)};
\addlegendentry{RTSNet False};

\addplot+[cyan!60!black, mark=o]
  coordinates {(-10,4.8) (0,-4.7) (10,-15.3) (20,-20) (30,-27.1)};
\addlegendentry{M-RTSNETP};

\addplot+[brown, solid, mark=diamond*]
  coordinates {(-10,2.9) (0,-6.3) (10,-13.9) (20,-16.8) (30,-21.1)};
\addlegendentry{BGRU};

\addplot+[cyan!60!black, dashed, mark=square*]
  coordinates {(-10,4.6) (0,-4.6) (10,-15.6) (20,-21.1) (30,-27.2)};
\addlegendentry{EM-KalmanNet};

\end{axis}
\end{tikzpicture}
\caption{MSE vs. SNR, linear Gaussian \ac{ss} model, Case 1.}
\label{fig:exp1_gauss}
\end{figure}
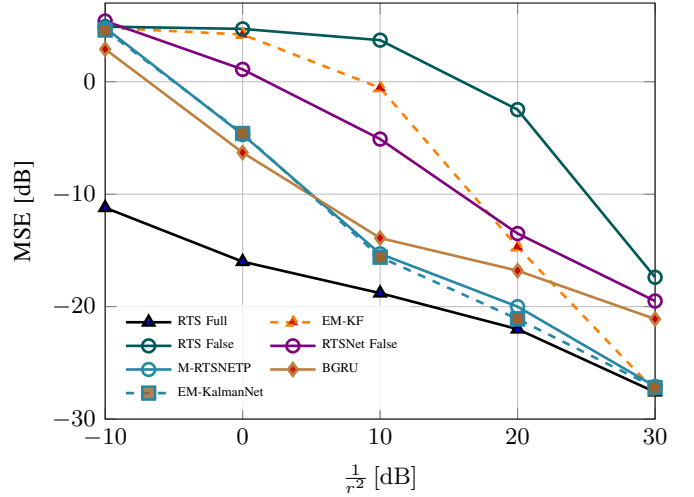

Fig.~\ref{fig:exp1_gauss} reports the state-estimation \ac{mse}. As expected, RTS Full provides the optimal benchmark, since it operates with the true block-specific
state-transition matrices under the linear Gaussian assumptions.

At low SNR, BGRU achieves the best performance among the non-oracle methods. In this regime, an MMSE optimal estimator relies more on the dynamical prior than on highly noisy observations, and BGRU's bidirectional recurrent architecture learns such temporal dependencies directly from the training data, without an explicit \ac{ss} model.
As SNR increases, EM-KalmanNet and M-RTSNETP increasingly outperform BGRU, since their model-based structure exploits the observation-state relation explicitly, which becomes more valuable as observations grow more informative.   
EM-KalmanNet closely tracks M-RTSNETP over the considered SNR range, with a small advantage at moderate-to-high SNR. The relatively narrow gap between the two methods is expected, since the analytical M-step employed by M-RTSNETP is well matched to the linear Gaussian setting, leaving limited room for the learned M-Net to improve upon it.

\subsubsection{Linear Non-Gaussian Setting}
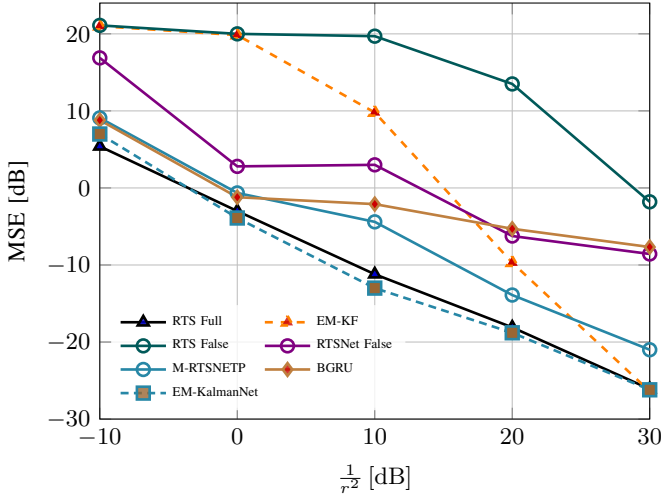
\begin{figure}
\centering
\begin{tikzpicture}
\begin{axis}[
    width=\linewidth,
    height=0.80\linewidth,
    xmin=-10, xmax=30,
    ymin=-30, ymax=24,
    xtick={-10,0,10,20,30},
    ytick={-30,-20,-10,0,10,20},
    grid=both,
    xlabel={$\frac{1}{r^{2}}\,[\mathrm{dB}]$},
    ylabel={MSE [dB]},
    label style={font=\small},
    tick label style={font=\small},
    every axis plot/.append style={line width=1.0pt, mark size=2.4pt},
    legend columns=2,
    clip=false,
    legend style={
        at={(0.02,0.02)},
        anchor=south west,
        draw=none,
        fill=white,
        fill opacity=0.80,
        text opacity=1,
        font=\tiny,
        rounded corners=1pt
    },
    legend cell align=left
]

\addplot+[black, mark=triangle*]
  coordinates {(-10,5.4) (0,-3) (10,-11.2) (20,-18.1) (30,-26.1)};
\addlegendentry{RTS Full};

\addplot+[orange, dashed, mark=triangle*]
  coordinates {(-10,21) (0,19.84) (10,9.8) (20,-9.7) (30,-26.4)};
\addlegendentry{EM-KF};

\addplot+[teal!70!black, mark=o]
  coordinates {(-10,21.1) (0,20) (10,19.7) (20,13.5) (30,-1.8)};
\addlegendentry{RTS False};

\addplot+[violet, mark=o]
  coordinates {(-10,16.88) (0,2.8) (10,3) (20,-6.24) (30,-8.57)};
\addlegendentry{RTSNet False};

\addplot+[cyan!60!black, mark=o]
  coordinates {(-10,9.09) (0,-0.65) (10,-4.4) (20,-13.9) (30,-21)};
\addlegendentry{M-RTSNETP};

\addplot+[brown, solid, mark=diamond*]
  coordinates {(-10,8.8) (0,-1.2) (10,-2.1) (20,-5.3) (30,-7.67)};
\addlegendentry{BGRU};

\addplot+[cyan!60!black, mark=square*]
  coordinates {(-10,7) (0,-3.9) (10,-13) (20,-18.8) (30,-26.2)};
\addlegendentry{EM-KalmanNet};

\end{axis}
\end{tikzpicture}
\caption{MSE vs. SNR, linear non-Gaussian \ac{ss} model, Case 1.}
\label{fig:exp2_exp}
\end{figure}

Fig.~\ref{fig:exp2_exp} reports the results for the linear model with exponentially distributed observation noise. Because the model-based smoothers' Gaussian assumption is now violated, RTS Full is no longer guaranteed to be \ac{mse}-optimal, and EM-KalmanNet closely tracks it throughout the tested range while consistently outperforming every other non-oracle method. 
The advantage over M-RTSNETP is substantially larger here than in
the Gaussian setting, which can be explained by the fact that M-RTSNETP's analytical M-step is derived under linear Gaussian assumptions, whereas M-Net of the EM-KalmanNet is free to adapt its parameter-update rule to the non-Gaussian statistics seen during training. Similarly, EM-KF degrades sharply when the observation noise is dominant, but recovers at high SNR, nearly matching the best-performing methods. As in the Gaussian experiment, BGRU improves more gradually as the SNR increases, widening its gap to the model-guided methods. These results demonstrate that EM-KalmanNet remains effective under non-Gaussian observation noise, addressing Challenge~\ref{itm:Noise}.

\subsubsection{Nonlinear Observation Setting}
\begin{figure}
\centering
\begin{tikzpicture}
\begin{axis}[
    width=\linewidth,
    height=0.80\linewidth,
    xmin=-10, xmax=30,
    ymin=-30, ymax=12,
    xtick={-10,0,10,20,30},
    ytick={-30,-20,-10,0,10},
    grid=both,
    xlabel={$\frac{1}{r^{2}}\,[\mathrm{dB}]$},
    ylabel={MSE [dB]},
    label style={font=\small},
    tick label style={font=\small},
    every axis plot/.append style={
        line width=1.0pt,
        mark size=2.4pt
    },
    legend columns=2,
    clip=false,
    legend style={
        at={(0.02,0.02)},
        anchor=south west,
        draw=none,
        fill=white,
        fill opacity=0.80,
        text opacity=1,
        font=\tiny,
        rounded corners=1pt
    },
    legend cell align=left,
    unbounded coords=jump
]

\addplot+[black, mark=triangle*]
coordinates {
    (-10,-10.9)
    (0,-16.5)
    (10,-19.2)
    (20,-22.2)
    (30,-26.9)
};
\addlegendentry{RTS Full};

\addplot+[orange, dashed, mark=triangle*]
coordinates {
    (-10,6.25)
    (0,5.5)
    (10,3.7)
    (20,-7.6)
    (30,-25)
};
\addlegendentry{EM-KF};

\addplot+[teal!70!black, mark=o]
coordinates {
    (-10,6)
    (0,5.2)
    (10,4.7)
    (20,-1.3)
    (30,-14.3)
};
\addlegendentry{RTS False};

\addplot+[violet, mark=o]
coordinates {
    (-10,5.02)
    (0,0.9)
    (10,-0.22)
    (20,-12.2)
    (30,-12)
};
\addlegendentry{RTSNet False};

\addplot+[cyan!60!black, solid, mark=o]
coordinates {
    (-10,3.56)
    (0,-2.14)
    (10,-12)
    (20,-21.3)
    (30,-22.3)
};
\addlegendentry{M-RTSNETP};

\addplot+[brown, solid, mark=diamond*]
coordinates {
    (-10,-2.9)
    (0,-9.5)
    (10,-14.7)
    (20,-15.7)
    (30,-20.6)
};
\addlegendentry{BGRU};

\addplot+[cyan!60!black, dashed, mark=square*]
coordinates {
    (-10,-1.2)
    (0,-8.67)
    (10,-16.7)
    (20,-19.9)
    (30,-24.6)
};
\addlegendentry{EM-KalmanNet};

\end{axis}
\end{tikzpicture}

\caption{MSE vs. SNR, nonlinear observation model, Case 1.}
\label{fig:exp3_non_linear_h}
\end{figure}
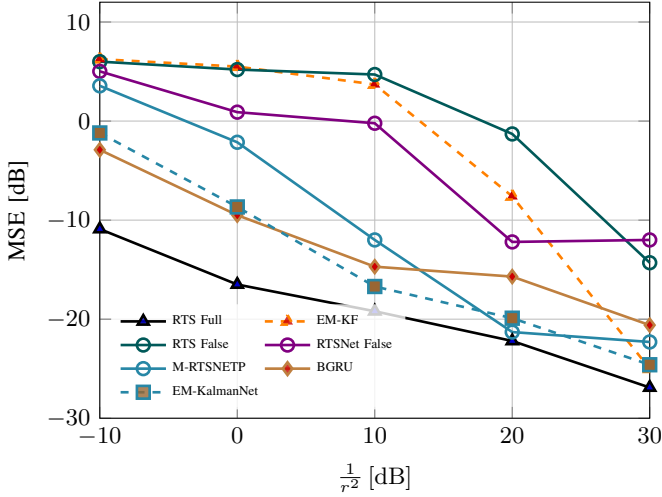

Fig.~\ref{fig:exp3_non_linear_h} reports the results for the nonlinear observation model, where RTS Full remains the strongest benchmark due to the accurate local linearization of $h(\cdot)$.
EM-KalmanNet is nonetheless among the strongest non-oracle methods across the tested \ac{snr} range: BGRU performs slightly better at the two lowest SNR levels, EM-KalmanNet leads clearly at moderate-to-high SNR, and at the highest SNR, EM-KF closely matches EM-KalmanNet while M-RTSNETP becomes competitive as well. Overall, EM-KalmanNet maintains strong performance under nonlinear observation models, addressing Challenge~\ref{itm:Approx}.


\begin{table}[t]
\centering

\caption{Inference latency,  nonlinear synthetic experiment, Case 1.}
\label{tab:synthetic_F_nonlinear_latency}
\begin{tabular}{l c}
\hline
Method & Latency [ms/sequence] \\
\hline
RTS & 190 \\
RTSNet & 180 \\
EM-KF, $3$ iterations & 600 \\
M-RTSNETP & 512 \\
BGRU &  0.23 \\
EM-KalmanNet & 335 \\
\hline
\end{tabular}
\end{table}

Table~\ref{tab:synthetic_F_nonlinear_latency} reports inference latency per sequence for the nonlinear observation experiment, addressing Challenge~\ref{itm:latency}. All timings are evaluated on an  NVIDIA GeForce RTX 4070 SUPER GPU platform using PyTorch, hence the notably lower latency of the BGRU, which is highly accelerated by PyTorch. Although EM-KF is restricted to only $3$ EM iterations, EM-KalmanNet remains considerably faster, reducing the latency from $600$ to $335$ [ms] ($\approx\!44.2\%$), and relative to M-RTSNETP from $512$ to $335$ [ms] ($\approx\!34.6\%$), while achieving comparable or better estimation accuracy across the tested SNR range (Figs.~\ref{fig:exp1_gauss}--\ref{fig:exp3_non_linear_h}). This gap between EM-KalmanNet and the iterative model-based baselines widens further in the more demanding settings considered later in this section.


\subsection{Synthetic Observation-Matrix Adaptation (Case 2)}
\label{subsec:synthetic_H}

\subsubsection{Setup}
The experimental setup follows that of Subsection~\ref{subsec:synthetic_F}, except that the unknown block-wise varying parameter is now $\myMat{H}$,
while the state-transition matrix $\myMat{F}_0$ remains fixed and known. The nominal
observation matrix $\myMat{H}_0$ is as in \eqref{eqn:Hmat}.
The observation matrix of the first block is obtained by rotating
$\myMat{H}_0$ by $0.2$ radians, and each subsequent block applies an
additional rotation of $0.2$ radians.
 We set $q^2=r^2$, such that the process noise is non-negligible relative to the measurement noise throughout; given the substantial performance degradation observed for all methods at $1/r^2=-10$ dB, we report here $1/r^2\in\{0,10,20,30\}$ dB.

\subsubsection{Linear Gaussian Setting}
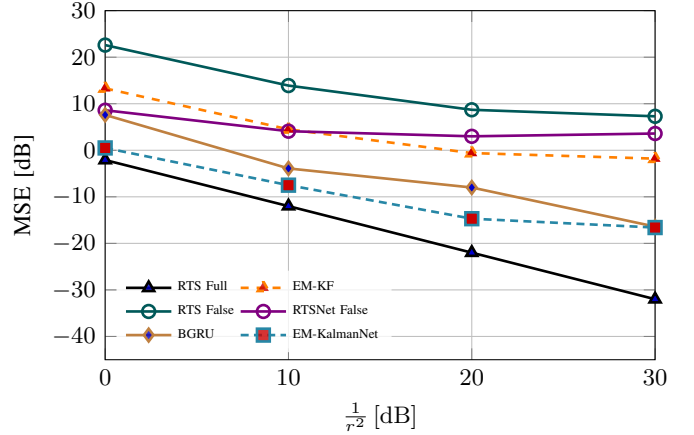
\begin{figure}
\centering
\begin{tikzpicture}
\begin{axis}[
    width=\linewidth,
    height=\figheight,
    xmin=0, xmax=30,
    ymin=-45, ymax=30,
    xtick={0,10,20,30},
    ytick={-40,-30,-20,-10,0,10,20,30},
    grid=both,
    xlabel={$\frac{1}{r^{2}}\,[\mathrm{dB}]$},
    ylabel={MSE [dB]},
    label style={font=\small},
    tick label style={font=\small},
    every axis plot/.append style={line width=1.0pt, mark size=2.4pt},
    legend columns=2,
    clip=false,
    legend style={
        at={(0.02,0.02)},
        anchor=south west,
        draw=none,
        fill=white,
        fill opacity=0.80,
        text opacity=1,
        font=\tiny,
        rounded corners=1pt
    },
    legend cell align=left,
    unbounded coords=jump
]

\addplot+[black, mark=triangle*]
  coordinates { (0,-2.1) (10,-12) (20,-22) (30,-32) };
\addlegendentry{RTS Full};

\addplot+[orange, dashed, mark=triangle*]
  coordinates { (0,13.3) (10,4.5) (20,-0.6) (30,-1.8) };
\addlegendentry{EM-KF};

\addplot+[teal!70!black, mark=o]
  coordinates { (0,22.6) (10,13.9) (20,8.7) (30,7.3) };
\addlegendentry{RTS False};

\addplot+[violet, mark=o]
  coordinates { (0,8.6) (10,4.1) (20,3) (30,3.6)};
\addlegendentry{RTSNet False};

\addplot+[brown, solid, mark=diamond*]
  coordinates { (0,7.6) (10,-3.9) (20,-8) (30,-16.4)};
\addlegendentry{BGRU};

\addplot+[cyan!60!black, mark=square*]
  coordinates { (0,0.5) (10,-7.5) (20,-14.7) (30,-16.6)};
\addlegendentry{EM-KalmanNet};

\end{axis}
\end{tikzpicture}
\caption{MSE vs. SNR,  linear Gaussian \ac{ss} model, Case 2.}
\label{fig:synthetic_H_gauss}
\end{figure}

Fig.~\ref{fig:synthetic_H_gauss} shows that, with RTS Full as the oracle benchmark, EM-KalmanNet generally achieves the strongest non-oracle \ac{mse} performance, and substantially outperforms both RTS False and RTSNet
False, demonstrating the benefit of adapting the observation matrix across blocks. It also consistently improves over EM-KF, while BGRU is competitive only at the lowest observation-noise variance.
These results demonstrate that the learned M-Net can effectively adapt the unknown observation matrix $\myMat{H}$, not merely $\myMat{F}$, 
%
and that this holds even when the process noise remains comparable to the observation noise throughout the tested range.

\subsubsection{Linear Non-Gaussian Setting}
\begin{figure}
\centering
\begin{tikzpicture}
\begin{axis}[
    width=\linewidth,
    height=\figheight,
    xmin=0, xmax=30,
    ymin=-35, ymax=45,
    xtick={0,10,20,30},
    ytick={-30,-20,-10,0,10,20,30,40},
    grid=both,
    xlabel={$\frac{1}{r^{2}}\,[\mathrm{dB}]$},
    ylabel={MSE [dB]},
    label style={font=\small},
    tick label style={font=\small},
    every axis plot/.append style={line width=1.0pt, mark size=2.4pt},
    legend columns=2,
    clip=false,
    legend style={
        at={(0.02,0.02)},
        anchor=south west,
        draw=none,
        fill=white,
        fill opacity=0.80,
        text opacity=1,
        font=\tiny,
        rounded corners=1pt
    },
    legend cell align=left,
    unbounded coords=jump
]

\addplot+[black, mark=triangle*]
  coordinates { (0,1) (10,-8.9) (20,-8.5) (30,-28.8) };
\addlegendentry{RTS Full};

\addplot+[orange, dashed, mark=triangle*]
  coordinates { (0,33.6) (10,23.9) (20,-1.2) (30,7.2) };
\addlegendentry{EM-KF};

\addplot+[teal!70!black, mark=o]
  coordinates { (0,42.4) (10,32.8) (20,5.5) (30,16) };
\addlegendentry{RTS False};

\addplot+[violet, mark=o]
  coordinates { (0,24.7) (10,12.9) (20,9.9) (30,7.95)};
\addlegendentry{RTSNet False};

\addplot+[brown, solid, mark=diamond*]
  coordinates { (0,14.5) (10,11.02) (20,-3.) (30,-11.4)};
\addlegendentry{BGRU};

\addplot+[cyan!60!black, mark=square*]
  coordinates { (0,16.4) (10,6.1) (20,-4.6) (30,-17.1)};
\addlegendentry{EM-KalmanNet};

\end{axis}
\end{tikzpicture}
\caption{MSE vs. SNR,  linear non-Gaussian \ac{ss} model, Case 2.}
\label{fig:synthetic_H_exp}
\end{figure}
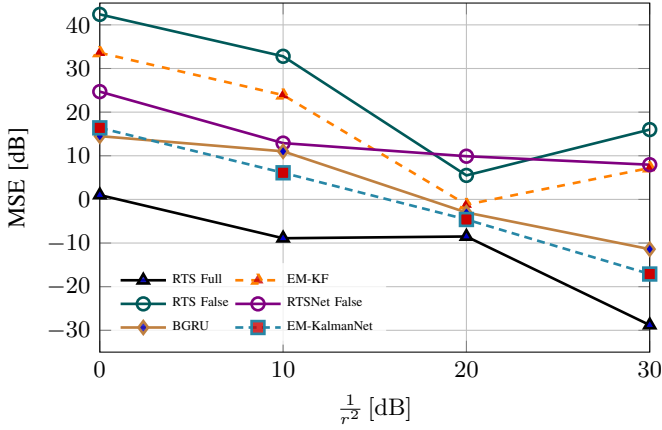

Fig.~\ref{fig:synthetic_H_exp} reports the results under exponential observation noise. RTS Full remains the strongest empirical benchmark, although it is no longer guaranteed to be MSE-optimal under the non-Gaussian noise distribution. EM-KalmanNet achieves the strongest non-oracle \ac{mse} performance over most of the tested observation-noise levels, with BGRU performing slightly better only at the lowest $1/r^2$. 
Together with the preceding Gaussian results, these results show that M-Net can effectively adapt the unknown block-varying observation matrix $\myMat{H}$ under non-Gaussian observation noise as well, thereby addressing  Challenge~\ref{itm:Noise}.

\subsection{Lorenz Attractor}
\label{subsec:lorenz}
The Lorenz attractor experiments \cite{lorenz1963deterministic} revisit each of the challenges introduced in Section~\ref{sec:system-model-and-preliminaries}: The chaotic Lorenz dynamics constitute a highly nonlinear, hard-to-approximate model (Challenge~\ref{itm:Approx}). Building on this model, we first evaluate adaptation to substantial changes in the observation matrix across datasets (Challenge~\ref{itm:varying}); we report the inference latency (Challenge~\ref{itm:latency}); we then evaluate robustness to different observation-noise distributions (Challenge~\ref{itm:Noise});  and finally, we evaluate operation across a range of block lengths (Challenge~\ref{itm:shortBlocks}).

\subsubsection{Setup}
The state evolves according to the discretized Lorenz attractor,  a three-dimensional chaotic solution of the Lorenz ordinary differential equations, leading to the standard Lorenz-63 state evolution (see \cite[Sec. IV-D]{revach2022kalmannet}). 
%
The observation model is linear in the state and is given by $
\myVec{y}_{t}=\myMat{H}_{i}\myVec{x}_{t}+\myVec{v}_{t}$, 
where $\myMat{H}_{i}$ is the observation matrix used in the $i$th block.  Throughout this subsection, the unknown  block-wise time-varying parameter is $\myMat{H}$, while the Lorenz state-transition function is fixed and known.

All compared methods are initialized with the nominal matrix
$\myMat{H}_{0}
=
\myMat{R}_{\mathrm{rot}}(\alpha_0)$, where $\myMat{R}_{\mathrm{rot}}(\cdot)$ is the three-dimensional rotation matrix and $
\alpha_0=1^{\circ}$. 
RTS False and RTSNet False retain
$\myMat{H}_{0}$ throughout all datasets;  the adaptive methods (EM-KF, EM-KalmanNet) are subsequently warm-started using their own estimate from the preceding dataset.


\subsubsection{Observation-Matrix Adaptation}

Here, the test sequence comprises five consecutive datasets, each of length $T=30$, with observation matrices generated by a fixed cumulative rotation, 
\begin{align}
\myMat{H}_{0}^{\mathrm{test}}
&=
\myMat{H}_{0},\\
\myMat{H}_{i}^{\mathrm{test}}
&=
\myMat{R}_{\mathrm{rot}}(0.3)
\myMat{H}_{i-1}^{\mathrm{test}},
\qquad i=1,\ldots,5,
\end{align}
applying the same $0.3$-radian angle about all three coordinate axes.
Training trajectories share this five-dataset structure, but each rotation angle is instead drawn independently with
\begin{align}
\myMat{H}_{0}^{(j)}
&=
\myMat{I}_3,\\
\myMat{H}_{i}^{(j)}
&=
\myMat{R}_{\mathrm{rot}}\big(\psi_{i}^{(j)}\big)
\myMat{H}_{i-1}^{(j)},
\qquad i=1,\ldots,5,
\end{align}
where the rotation angle is independently sampled for each dataset and
trajectory from 
$\mathcal{U}[-0.3,0.3
]$ radians for the three coordinate axes. 
%
%
%
The test setting is particularly challenging: the true rotation is applied cumulatively at the boundary of the training range, so the test trajectories exhibit a systematic drift only sparsely represented during training. Strong performance here indicates that EM-KalmanNet learns a transferable adaptation strategy rather than memorizing training trajectories.

\begin{figure}
\centering
\begin{tikzpicture}
\begin{axis}[
    width=\linewidth,
    height=\figheight,
    xmin=-10, xmax=30,
    ymin=-45, ymax=32,
    xtick={-10,0,10,20,30},
    ytick={-40,-30,-20,-10,0,10,20,30},
    grid=both,
    xlabel={$\frac{1}{r^{2}}\,[\mathrm{dB}]$},
    ylabel={MSE [dB]},
    label style={font=\small},
    tick label style={font=\small},
    every axis plot/.append style={line width=1.0pt, mark size=2.4pt},
    legend columns=2,
    clip=false,
    legend style={
        at={(0.02,0.02)},
        anchor=south west,
        draw=none,
        fill=white,
        fill opacity=0.80,
        text opacity=1,
        font=\tiny,
        rounded corners=1pt
    },
    legend cell align=left,
    unbounded coords=jump
]

\addplot+[black, mark=triangle*]
  coordinates { (-10,-2.5) (0,-12.45) (10,-22.3) (20,-33.2) (30,-42.26)};
\addlegendentry{RTS Full};

\addplot+[orange, dashed, mark=triangle*]
  coordinates { (-10,6.3) (0,3.2) (10,2.2) (20,2) (30,2)};
\addlegendentry{EM-KF};

\addplot+[teal!70!black, mark=o]
  coordinates { (-10,20.2) (0,21.5) (10,20.9) (20,20.95) (30,21.5)};
\addlegendentry{RTS False};

\addplot+[brown, mark=diamond*]
  coordinates {(-10,7.8) (0,4.8) (10,5.5) (20,9) (30,9.8)};
\addlegendentry{BGRU};

\addplot+[violet, mark=o]
  coordinates { (-10,9.4) (0,0.97) (10,0.78) (20,0) (30,-1.7)};
\addlegendentry{RTSNet False};

\addplot+[cyan!60!black, mark=square*]
  coordinates { (-10,1.45) (0,-4.9) (10,-6.8) (20,-8.4) (30,-10.8)};
\addlegendentry{EM-KalmanNet};

\end{axis}
\end{tikzpicture}
\caption{MSE vs. SNR,  Lorenz attractor, Case 2.}
\label{fig:lorenz_5datasets}
\end{figure}
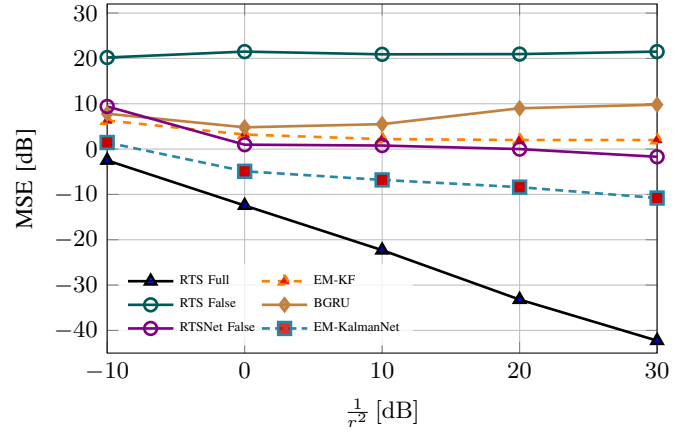

Fig.~\ref{fig:lorenz_5datasets} shows this mismatch strongly degrades BGRU, whose fully data-driven operation depends heavily on the similarity between the training and test trajectories. EM-KalmanNet is more robust, as it combines learned inference with explicit \ac{ss} structure and block-wise adaptation of $\myMat{H}$, and therefore outperforms all non-oracle methods. It also improves over EM-KF, which struggles to recover the large accumulated changes within the limited number of iterations.
Due to the substantial cumulative variation in the observation model,
none of the adaptive methods approaches the oracle RTS Full, which receives the true $\myMat{H}_i$  in every dataset. Nevertheless, EM-KalmanNet is the strongest and most consistent among the non-oracle methods.

 Table~\ref{tab:lorenz_latency} reports the latency per complete five-block sequence for this nonlinear, chaotic setting. Despite its $3$-iteration budget, EM-KalmanNet reduces the latency relative to EM-KF from $868$ to $242$ [ms] ($\approx\!72.1\%$); RTSNet False and BGRU are faster still, but at substantially inferior estimation accuracy.

\begin{table}
\centering 
\begin{tabular}{l c} 
\hline Method & Latency [ms/sequence] \\
\hline RTS Full & 280 \\
EM-KF, $3$ iterations & 868 \\
BGRU & 0.4 \\
RTSNet False & 83 \\
EM-KalmanNet & 242 \\
\hline 
\end{tabular}
\caption{Inference latency, Lorenz attractor, $T=30$.} \label{tab:lorenz_latency}
\end{table}

\subsubsection{Robustness Across Observation-Noise Distributions}
We next repeat the preceding Lorenz experiment under six different observation-noise distributions, keeping the remaining setup unchanged, to evaluate robustness beyond the Gaussian noise assumption (Challenge~\ref{itm:Noise}). 
 Alongside standard Gaussian noise, we consider a correlated Gaussian setting. 
For the correlated Gaussian setting, the observation-noise covariance is
\begin{align}
\myMat{R}
=
r^2
\begin{bmatrix}
1 & 0.6 & 0.6\\
0.6 & 1 & 0.6\\
0.6 & 0.6 & 1
\end{bmatrix},\nonumber
\end{align}
such that we have equal per-channel variance $r^2$ and pairwise correlation $\rho=0.6$.
In addition, we have symmetric, heavy-tailed Laplace noise; Student-$t$ noise with $\nu=4$ degrees of freedom, producing frequent large-amplitude deviations; exponential noise, centered to zero mean while retaining its asymmetric, skewed shape; and contaminated Gaussian noise, modeling a predominantly Gaussian process with rare high-amplitude outliers ($\epsilon=0.05$, $\kappa=3$). These settings jointly probe cross-channel correlation, tail heaviness, asymmetry, and impulsiveness. 


Table~\ref{tab:lorenz_noise_robustness} shows that EM-KalmanNet achieves the lowest MSE among the non-oracle methods across all tested noise settings, with a particularly clear margin under the heavy-tailed Laplace and contaminated Gaussian distributions. The correlated Gaussian case is hardest for all methods, as the shared cross-channel noise component reduces the independent information available in the observations, yet EM-KalmanNet's advantage persists. Its standard deviations are also consistently smaller than BGRU's, indicating more reliable performance across test realizations rather than a strong dependence on specific temporal patterns seen in training.
Overall, these results demonstrate the robustness
of EM-KalmanNet to diverse noise statistics, addressing
Challenge~\ref{itm:Noise}.

\begin{table}
\centering
\renewcommand{\arraystretch}{1.35}
\setlength{\tabcolsep}{4pt}
\scriptsize

\resizebox{\columnwidth}{!}{%
\begin{tabular}{|c|c|c|c|c|c|c|}
\hline

\textbf{Method}
& \textbf{Gaussian}
& \shortstack[c]{\textbf{Correlated}\\\textbf{Gaussian}}
& \textbf{Laplace}
& \textbf{Exponential}
& \textbf{Student-$t$}
& \textbf{Contaminated}
\\
\hline

RTS Full
& $-12.43$
& $-12.90$
& $-12.30$
& $-12.36$
& $-10.20$
& $-11.50$
\\[-1pt]

&
$\pm 2.50$
& $\pm 2.47$
& $\pm 2.45$
& $\pm 2.48$
& $\pm 2.80$
& $\pm 2.50$
\\[2pt]
\hline

RTSNet False
& $-0.97$
& $1.08$
& $-0.90$
& $-1.80$
& $-1.20$
& $-1.20$
\\[-1pt]

&
$\pm 4.70$
& $\pm 5.10$
& $\pm 5.00$
& $\pm 4.37$
& $\pm 4.39$
& $\pm 4.50$
\\[2pt]
\hline

EM-KF
& $2.50$
& $7.59$
& $2.75$
& $2.71$
& $2.88$
& $3.47$
\\[-1pt]

&
$\pm 4.60$
& $\pm 3.66$
& $\pm 4.39$
& $\pm 4.50$
& $\pm 4.33$
& $\pm 4.47$
\\[2pt]
\hline

BGRU
& $4.80$
& $7.96$
& $6.30$
& $5.50$
& $6.70$
& $3.28$
\\[-1pt]

&
$\pm 8.70$
& $\pm 8.57$
& $\pm 9.17$
& $\pm 8.85$
& $\pm 8.98$
& $\pm 7.50$
\\[2pt]
\hline

EM-KalmanNet
& $-4.90$
& $-3.36$
& $-6.30$
& $-4.20$
& $-3.86$
& $-6.20$
\\[-1pt]

&
$\pm 4.60$
& $\pm 5.48$
& $\pm 4.10$
& $\pm 4.82$
& $\pm 4.34$
& $\pm 4.00$
\\[2pt]
\hline

\end{tabular}%
}
\caption{State-estimation MSE (mean $\pm$ std) under
different observation-noise distributions, Lorenz attractor.}
\label{tab:lorenz_noise_robustness}
\end{table}


\subsubsection{Effect of the Block Length}
We next vary $T$ to evaluate adaptation to the changing observation model
when only a limited number of observations is available, addressing
Challenge~\ref{itm:shortBlocks}. We set $r^2=1$ and $q^2=0.01$, and
consider three consecutive blocks with a cumulative rotation of $0.2$
radians between them. We additionally report EM-KF with $10$ iterations
to assess whether a larger iterative budget can compensate for short
blocks.

\begin{figure}
\centering
\begin{tikzpicture}
\begin{axis}[
    width=\linewidth,
    height=0.78\linewidth,
    xmin=10, xmax=50,
    ymin=-15, ymax=30,
    xtick={10,20,30,40,50},
    ytick={-15,-10,-5,0,5,10,15,20,25},
    grid=both,
    xlabel={Block length $T$},
    ylabel={MSE [dB]},
    label style={font=\small},
    tick label style={font=\small},
    every axis plot/.append style={line width=1.0pt, mark size=2.4pt},
    legend columns=3,
    clip=false,
    legend style={
        at={(0.5,0.98)},
        anchor=north,
        draw=none,
        fill=white,
        fill opacity=0.80,
        text opacity=1,
        font=\tiny,
        rounded corners=1pt
    },
    legend cell align=left,
    unbounded coords=jump
]

\addplot+[black, mark=triangle*]
  coordinates {(10,-12.56) (20,-13) (30,-13.3) (40,-13.54) (50,-13.6)};
\addlegendentry{RTS Full};

\addplot+[orange, dashed, mark=triangle*]
  coordinates {(10,10.9) (20,-2.5) (30,-7.8) (40,-9.5) (50,-10.3)};
\addlegendentry{EM-KF 10 iterations};

\addplot+[orange, dashed, mark=o]
  coordinates {(10,27.9) (20,6.7) (30,-1.56) (40,-5.3) (50,-5.3)};
\addlegendentry{EM-KF 3 iterations};

\addplot+[teal!70!black, mark=o]
  coordinates {(10,12.25) (20,6.7) (30,9.6) (40,8.4) (50,9.3)};
\addlegendentry{RTS False};

\addplot+[violet, mark=o]
  coordinates {(10,-0.3) (20,-7.2) (30,-4) (40,-2.4) (50,-0.6)};
\addlegendentry{RTSNet False};

\addplot+[cyan!60!black, mark=square*]
  coordinates {(10,-4) (20,-8.1) (30,-8.7) (40,-9.6) (50,-10.32)};
\addlegendentry{EM-KalmanNet};

\addplot+[brown, mark=diamond*]
  coordinates {(10,3.66) (20,-6.8) (30,-8.7) (40,-7.7) (50,-3.6)};
\addlegendentry{BGRU};

\end{axis}
\end{tikzpicture}
\caption{MSE vs. $T$, Lorenz attractor, $r^2=1$, $q^2=0.01$.}
\label{fig:lorenz_T_sweep}
\end{figure}
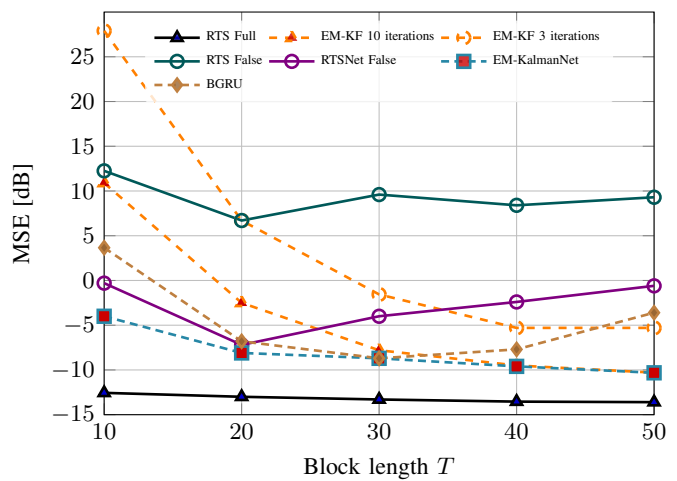

Fig.~\ref{fig:lorenz_T_sweep} reports the \ac{mse} versus different block lengths. RTS Full provides the strongest performance, whereas RTS False remains severely degraded by the fixed model 
mismatch. EM-KalmanNet maintains the best non-oracle performance across all tested $T$, with its advantage most pronounced for  $T<30$.
Even the $10$-iteration EM-KF variant cannot close this gap at short $T$, despite its added latency; EM-KF only approaches EM-KalmanNet as $T$ grows and more observations become available.
BGRU is competitive for small $T$, where short-range temporal patterns are easier to capture from the training data, but its performance degrades as $T$ increases.
RTSNet False is similarly competitive for short blocks but, lacking explicit adaptation of $\myMat{H}$, benefits less from longer blocks, leading to a growing gap to EM-KalmanNet.
These results demonstrate that EM-KalmanNet can adapt 
from a limited number of observations,
 addressing~\ref{itm:shortBlocks}.

\subsubsection{Stability Over an Extended Horizon}
We next evaluate repeated adaptation over ten consecutive datasets of length $T=30$, with $\myMat{H}$ fixed within each dataset and cumulatively rotated by $\psi=0.1$ radians between datasets, using a harder noise level $r^2=10$, $q^2=0.1$.
 Unlike the five-dataset experiment above,  which considers larger changes over fewer datasets, this setting applies smaller but more frequent changes over a longer horizon. The goal is to assess whether successive parameter updates remain stable and whether estimation errors accumulate over time.

\begin{figure*}[!t]
\centering

\newcommand{\lorenzimage}[1]{%
    \includegraphics[
        width=\linewidth,
        trim=2mm 3mm 3mm 20mm,
        clip
    ]{#1}%
}


\begin{minipage}[t]{0.235\textwidth}
    \centering
    \lorenzimage{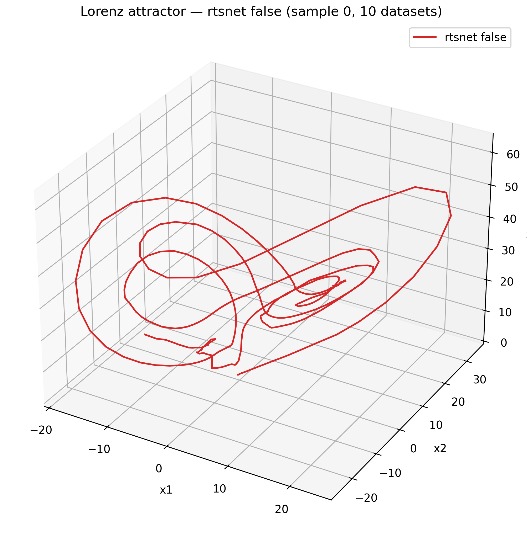}

    \vspace{1mm}
    {\small RTSNet False}
\end{minipage}
\hfill
\begin{minipage}[t]{0.235\textwidth}
    \centering
    \lorenzimage{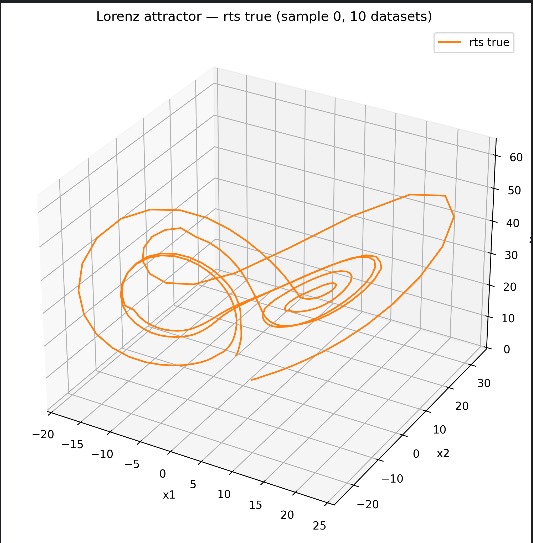}

    \vspace{1mm}
    {\small RTS Full}
\end{minipage}
\hfill
\begin{minipage}[t]{0.235\textwidth}
    \centering
    \lorenzimage{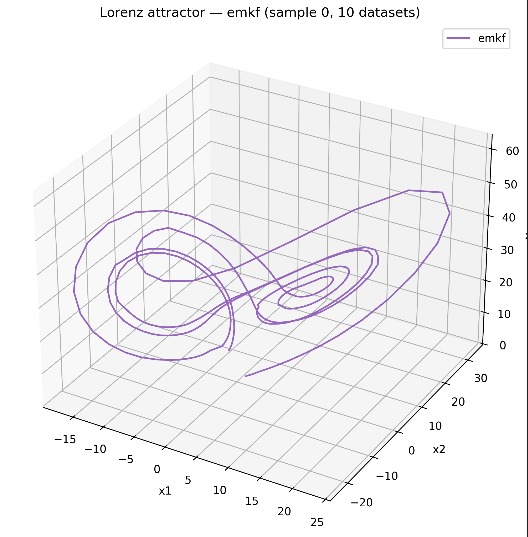}

    \vspace{1mm}
    {\small EM-KF}
\end{minipage}
\hfill
\begin{minipage}[t]{0.235\textwidth}
    \centering
    \lorenzimage{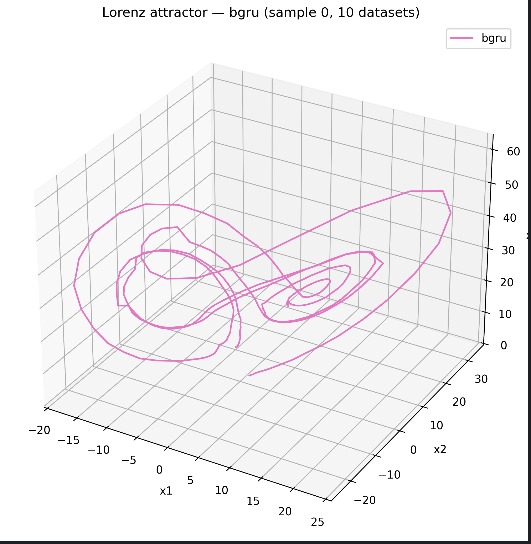}

    \vspace{1mm}
    {\small BGRU}
\end{minipage}

\vspace{4mm}


\begin{minipage}[t]{0.235\textwidth}
    \centering
    \lorenzimage{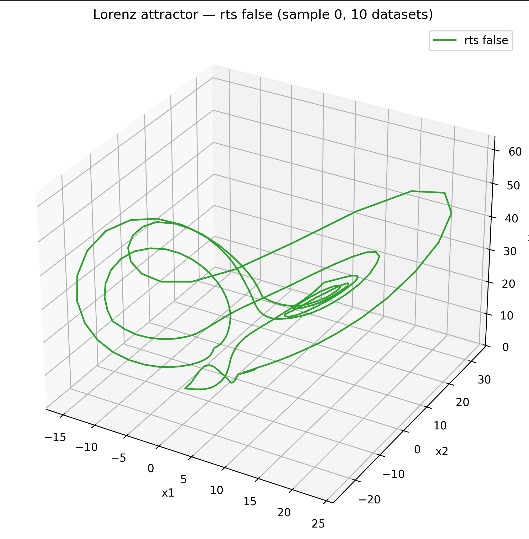}

    \vspace{1mm}
    {\small RTS False}
\end{minipage}
\hfill
\begin{minipage}[t]{0.235\textwidth}
    \centering
    \lorenzimage{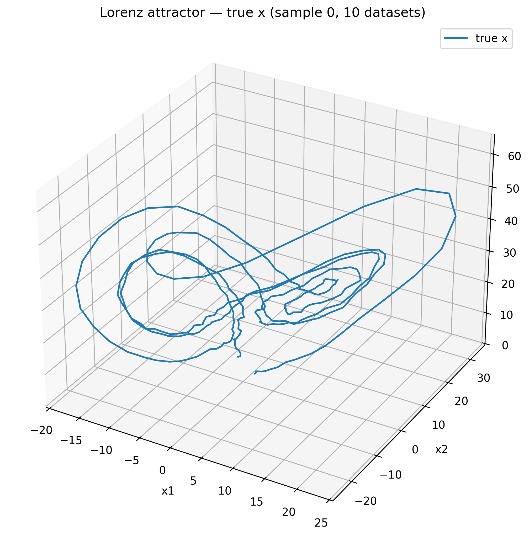}

    \vspace{1mm}
    {\small Ground Truth}
\end{minipage}
\hfill
\begin{minipage}[t]{0.235\textwidth}
    \centering
    \lorenzimage{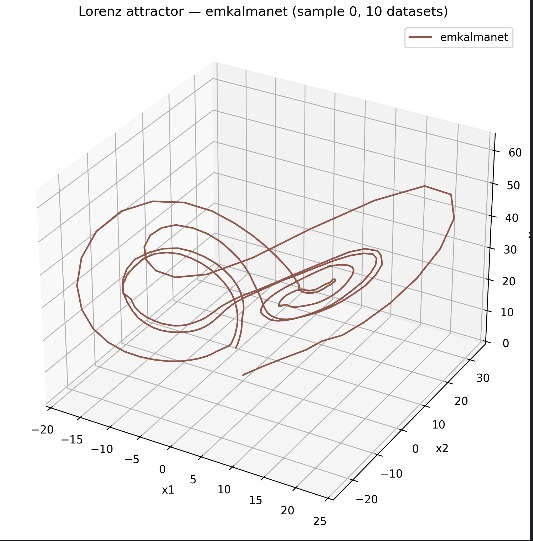}

    \vspace{1mm}
    {\small EM-KalmanNet}
\end{minipage}

\caption{Representative state-estimation trajectories for the Lorenz
attractor experiment with ten consecutive datasets,
and a cumulative observation-model rotation of $\psi=0.1$ radians
between successive datasets. }
\label{fig:lorenz_10datasets}
\end{figure*}

Table~\ref{tab:lorenz_10datasets} shows that EM-KalmanNet achieves the
strongest performance among the non-oracle methods.
These results indicate that the repeated M-Net updates remain effective
over the extended sequence without substantial error accumulation.
Moreover, EM-KalmanNet exhibits a relatively low standard deviation of
$2.6$ dB, close to that of the oracle RTS Full benchmark. 
Fig.~\ref{fig:lorenz_10datasets} further illustrates how these numerical
differences translate into the reconstructed state trajectories.
EM-KalmanNet closely preserves the characteristic geometry of the
Lorenz attractor and follows the ground-truth trajectory more faithfully
than the other non-oracle methods.

\begin{table}
\centering

\renewcommand{\arraystretch}{1.35}
\setlength{\tabcolsep}{3.5pt}
\scriptsize

\resizebox{\columnwidth}{!}{%
\begin{tabular}{|c|c|c|c|c|c|}
\hline

 \textbf{RTS Full}
& \textbf{RTS False}
& \textbf{RTSNet False}
& \textbf{EM-KF}
& \textbf{BGRU}
& \textbf{EM-KalmanNet}
\\
\hline

 $-3.3$
& $14.8$
& $12.1$
& $4.4$
& $1.7$
& $-0.86$
\\[-1pt]

$\pm\,\text{2.3}$
& $\pm\,\text{7.4}$
& $\pm\,\text{7.6}$
& $\pm\,\text{4}$
& $\pm\,\text{3.48}$
& $\pm\,\text{2.6}$
\\[2pt]

\hline
\end{tabular}%
}
\caption{State-estimation MSE (mean $\pm$ standard deviation) for the
Lorenz attractor experiment with ten consecutive datasets, $T=30$,
$r^2=10$, and an observation-model rotation of $\psi=0.1$ radians
between successive datasets. 
}
\label{tab:lorenz_10datasets}
\end{table}


\subsection{Acoustic Tracking}
\label{subsec:acoustic}

We finally consider a real-world-inspired acoustic tracking scenario,
in which a moving sound source with varying motion dynamics is localized
from measurements collected by a microphone array. This experiment
represents a practical setting where the source motion is not accurately
known a priori and the observations are nonlinear, allowing us to
evaluate the applicability of EM-KalmanNet beyond the preceding
synthetic setups~\cite{GannotDvorkind2006}. 

\subsubsection{Setup}
The hidden state consists of the two-dimensional source position and
velocity,
\begin{align}
\myVec{x}_t =
\begin{bmatrix}
p_{x,t} & p_{y,t} & v_{x,t} & v_{y,t}
\end{bmatrix}^{\top}.
\end{align}
The state evolves according to $
\myVec{x}_t
=
\myMat{F}(\psi)\myVec{x}_{t-1}
+
\myVec{w}_t$, 
where
\begin{align}
\myMat{F}(\psi)
=
\begin{bmatrix}
1 & 0 & \Delta t & 0\\
0 & 1 & 0 & \Delta t\\
0 & 0 & \cos\psi & -\sin\psi\\
0 & 0 & \sin\psi & \cos\psi
\end{bmatrix},
\qquad
\Delta t=0.1.
\end{align}
Thus, the position follows constant-velocity kinematics, while the
velocity vector is rotated by an angle $\psi$. The structure of
$\myMat{F}(\psi)$ is known, whereas $\psi$, and consequently the
block-specific state-transition matrix, is unknown.

The observations are time-difference-of-arrival  measurements
relative to a reference microphone,
\begin{align}
y_t^{(i)}
=
\frac{
\left\|\myVec{p}_t-\myVec{s}_i\right\|_2
-
\left\|\myVec{p}_t-\myVec{s}_{\rm ref}\right\|_2
}{c}
+
v_t^{(i)},
\end{align}
where $\myVec{p}_t=[p_{x,t},p_{y,t}]^{\top}$ and $c$ denotes the
propagation velocity. Hence, the observations are nonlinear functions
of the source position. The process- and observation-noise levels are
set to $q^2=0.01$ and $r^2=1$, respectively.

Each trajectory comprises six temporally consecutive blocks. The final
state of each block initializes the following block, while the motion
parameter $\psi$ remains fixed within a block and changes between
blocks. During training, $\psi$ is independently sampled for each
block and trajectory according to
$\psi\sim\mathcal{U}[-0.12,0.12]$. During testing, the six blocks use
the fixed sequence $
\psi =
[0.1,\;0.08,\;-0.1,\;-0.08,\;0.06,\;0.1]$.
The mismatched baselines assume $\psi=0$.
As in the previous experiments, EM-KalmanNet is initialized with
$\myMat{F}(0)$ at each block and estimates the block-specific
transition matrix without warm-starting from the preceding block.

\subsubsection{Results}
\begin{figure}
\centering
\includegraphics[width=0.85\linewidth]{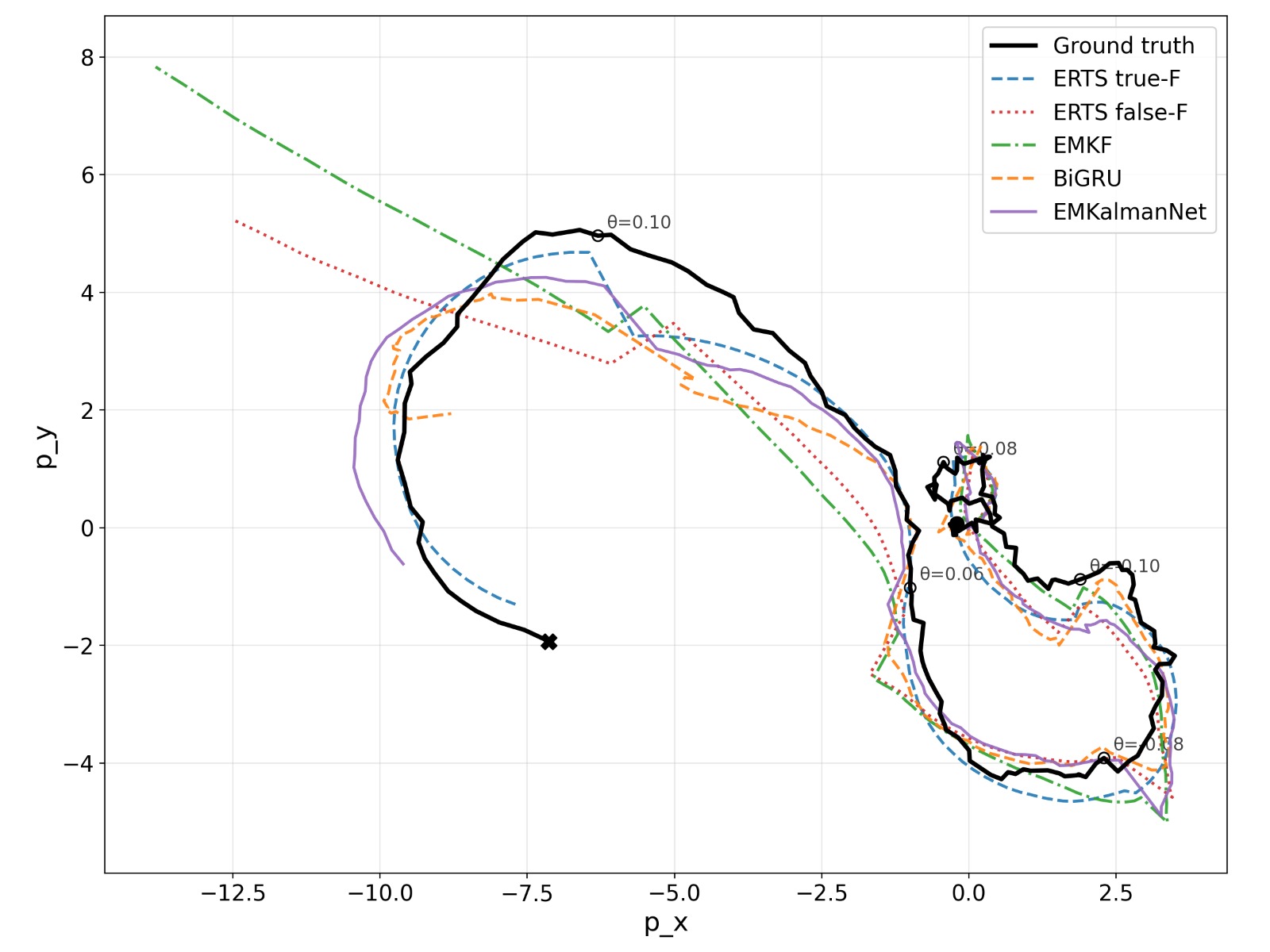}
\vspace{-0.2cm}
\caption{Acoustic tracking experiment with three microphones. }
\label{fig:acoustic_tracking}
\end{figure}

Fig.~\ref{fig:acoustic_tracking} illustrates the acoustic tracking
experiment, while Table~\ref{tab:acoustic_results} reports the overall
state-estimation \ac{mse}. As expected, ERTS with the true
block-specific transition matrix provides the strongest oracle
benchmark. In contrast, ERTS and RTSNet operating with the fixed
mismatched model $\psi=0$ degrade when the actual source motion
departs from the assumed constant-velocity dynamics.
EM-KalmanNet substantially improves upon these mismatched baselines by
adapting the state-transition matrix to the motion in each block.
Notably, this adaptation is performed from the same mismatched
initialization at the beginning of every block, demonstrating the
ability of EM-KalmanNet to repeatedly recover changing motion dynamics
without relying on the parameter estimate obtained in the preceding
block.

\begin{table}[t]
\centering
\begin{tabular}{l c}
\hline
Method & MSE [dB] \\
\hline
RTS True & $-8.39$ \\
RTS False & $-1.95$  \\
RTSNet False-$\myMat{F}$ & $-4.48$ \\
BGRU & $-4.73$ \\
EM-KalmanNet & $-6.1$ \\
\hline
\end{tabular}
\caption{Overall MSE for the acoustic tracking experiment.}
\label{tab:acoustic_results}
\end{table}

\section{Conclusion}
\label{sec:conc}

We considered state estimation in partially known state-space models
with block-wise time-varying parameters. We proposed EM-KalmanNet, which combines an RTSNet-based
learned E-step with a data-driven M-step for iterative state smoothing
and parameter adaptation. The numerical study shows that EM-KalmanNet
consistently improves over mismatched model-based and data-driven
benchmarks across linear and nonlinear dynamics, non-Gaussian noise,
and rapidly varying model conditions, while maintaining a fixed and
practical inference budget. 

\vspace{-0.25cm}
\bibliographystyle{IEEEtran}

\begin{thebibliography}{10}
\providecommand{\url}[1]{#1}
\csname url@samestyle\endcsname
\providecommand{\newblock}{\relax}
\providecommand{\bibinfo}[2]{#2}
\providecommand{\BIBentrySTDinterwordspacing}{\spaceskip=0pt\relax}
\providecommand{\BIBentryALTinterwordstretchfactor}{4}
\providecommand{\BIBentryALTinterwordspacing}{\spaceskip=\fontdimen2\font plus
\BIBentryALTinterwordstretchfactor\fontdimen3\font minus \fontdimen4\font\relax}
\providecommand{\BIBforeignlanguage}[2]{{%
\expandafter\ifx\csname l@#1\endcsname\relax
\typeout{** WARNING: IEEEtran.bst: No hyphenation pattern has been}%
\typeout{** loaded for the language `#1'. Using the pattern for}%
\typeout{** the default language instead.}%
\else
\language=\csname l@#1\endcsname
\fi
#2}}
\providecommand{\BIBdecl}{\relax}
\BIBdecl

\bibitem{Cohen2026EMKalmanNet}
O.~Cohen, X.~Ni, N.~Shlezinger, and T.~Routtenberg, ``{EM-K}almannet: {AI}-aided {K}alman tracking in partially known time-varying state-space models,'' in \emph{IEEE Sensor Array and Multichannel Signal Processing Workshop (SAM)}, 2026.

\bibitem{sarkka2023bayesian}
S.~S{\"a}rkk{\"a} and L.~Svensson, \emph{Bayesian filtering and smoothing}.\hskip 1em plus 0.5em minus 0.4em\relax Cambridge university press, 2023, vol.~17.

\bibitem{karami2020smart}
Z.~Karami and R.~Kashef, ``Smart transportation planning: Data, models, and algorithms,'' \emph{Transportation Engineering}, vol.~2, p. 100013, 2020.

\bibitem{bar2004estimation}
Y.~Bar-Shalom, X.~R. Li, and T.~Kirubarajan, \emph{Estimation with applications to tracking and navigation: theory algorithms and software}.\hskip 1em plus 0.5em minus 0.4em\relax John Wiley \& Sons, 2004.

\bibitem{durbin2012time}
J.~Durbin and S.~J. Koopman, \emph{Time series analysis by state space methods}.\hskip 1em plus 0.5em minus 0.4em\relax Oxford University Press, 2012.

\bibitem{kalman1960new}
R.~E. Kalman, ``A new approach to linear filtering and prediction problems,'' \emph{Journal of Basic Engineering}, vol.~82, no.~1, pp. 35--45, 1960.

\bibitem{rauch1965maximum}
H.~E. Rauch, F.~Tung, and C.~T. Striebel, ``Maximum likelihood estimates of linear dynamic systems,'' \emph{AIAA Journal}, vol.~3, no.~8, pp. 1445--1450, 1965.

\bibitem{schmidt1981kalman}
S.~F. Schmidt, ``The {K}alman filter-its recognition and development for aerospace applications,'' \emph{Journal of Guidance and Control}, vol.~4, no.~1, pp. 4--7, 1981.

\bibitem{julier2004unscented}
S.~J. Julier and J.~K. Uhlmann, ``Unscented filtering and nonlinear estimation,'' \emph{Proc. {IEEE}}, vol.~92, no.~3, pp. 401--422, 2004.

\bibitem{arasaratnam2009cubature}
I.~Arasaratnam and S.~Haykin, ``Cubature {K}alman filters,'' \emph{{IEEE} Trans. Autom. Control}, vol.~54, pp. 1254--1269, 2009.

\bibitem{mehra1970identification}
R.~Mehra, ``On the identification of variances and adaptive {K}alman filtering,'' \emph{{IEEE} Trans. Autom. Control}, vol.~15, no.~2, pp. 175--184, 1970.

\bibitem{blom1988interacting}
H.~A. Blom and Y.~Bar-Shalom, ``The interacting multiple model algorithm for systems with markovian switching coefficients,'' \emph{{IEEE} Trans. Autom. Control}, vol.~33, no.~8, pp. 780--783, 1988.

\bibitem{wan2001dual}
E.~A. Wan and A.~T. Nelson, ``Dual extended {K}alman filter methods,'' \emph{Kalman filtering and neural networks}, pp. 123--173, 2001.

\bibitem{Togneri2003}
R.~Togneri and L.~Deng, ``Joint state and parameter estimation for a target-directed nonlinear dynamic system model,'' \emph{{IEEE} Trans. Signal Process.}, vol.~51, no.~12, pp. 3061--3070, 2003.

\bibitem{zhang2020identification}
L.~Zhang, D.~Sidoti, A.~Bienkowski, K.~R. Pattipati, Y.~Bar-Shalom, and D.~L. Kleinman, ``On the identification of noise covariances and adaptive {K}alman filtering: A new look at a 50 year-old problem,'' \emph{{IEEE} Access}, vol.~8, pp. 59\,362--59\,388, 2020.

\bibitem{moon1996expectation}
T.~K. Moon, ``The expectation-maximization algorithm,'' \emph{{IEEE} Signal Process. Mag.}, vol.~13, no.~6, pp. 47--60, 1996.

\bibitem{gannot2008kalman}
S.~Gannot and A.~Yeredor, ``The {K}alman filter,'' \emph{Springer Handbook of Speech Processing}, pp. 135--160, 2008.

\bibitem{Shumway1982}
R.~H. Shumway and D.~S. Stoffer, ``An approach to time series smoothing and forecasting using the {EM} algorithm,'' \emph{Journal of Time Series Analysis}, vol.~3, pp. 253--264, 1982.

\bibitem{shlezinger2025artificial}
N.~Shlezinger \emph{et~al.}, ``Artificial intelligence-aided {K}alman filters: {AI}-augmented designs for {K}alman-type algorithms,'' \emph{{IEEE} Signal Process. Mag.}, vol.~42, no.~3, pp. 52--76, 2025.

\bibitem{haarnoja2016backprop}
T.~Haarnoja, A.~Ajay, S.~Levine, and P.~Abbeel, ``Backprop {KF}: Learning discriminative deterministic state estimators,'' \emph{Advances in Neural Information Processing Systems}, vol.~29, 2016.

\bibitem{becker2019recurrent}
P.~Becker, H.~Pandya, G.~Gebhardt, C.~Zhao, C.~J. Taylor, and G.~Neumann, ``Recurrent {K}alman networks: Factorized inference in high-dimensional deep feature spaces,'' in \emph{International Conference on Machine Learning}, 2019, pp. 544--552.

\bibitem{ghosh2023danse}
A.~Ghosh, A.~Honor{\'e}, and S.~Chatterjee, ``{DANSE}: Data-driven non-linear state estimation of model-free process in unsupervised learning setup,'' \emph{{IEEE} Trans. Signal Process.}, vol.~72, pp. 1824--1838, 2024.

\bibitem{satorras2019combining}
V.~G. Satorras, Z.~Akata, and M.~Welling, ``Combining generative and discriminative models for hybrid inference,'' \emph{Advances in Neural Information Processing Systems}, 2019.

\bibitem{klushyn2021latent}
A.~Klushyn \emph{et~al.}, ``Latent matters: Learning deep state-space models,'' \emph{Advances in Neural Information Processing Systems}, 2021.

\bibitem{shlezinger2020model}
N.~Shlezinger, J.~Whang, Y.~C. Eldar, and A.~G. Dimakis, ``Model-based deep learning,'' \emph{Proc. {IEEE}}, vol. 111, no.~5, pp. 465--499, 2023.

\bibitem{Xu2024EKFNet}
L.~Xu and R.~Niu, ``{EKFNet}: Learning system noise covariance parameters for nonlinear tracking,'' \emph{{IEEE} Trans. Signal Process.}, vol.~72, pp. 3139--3152, 2024.

\bibitem{revach2022kalmannet}
G.~Revach \emph{et~al.}, ``Kalman{N}et: Neural network aided {K}alman filtering for partially known dynamics,'' \emph{{IEEE} Trans. Signal Process.}, vol.~70, pp. 1532--1547, 2022.

\bibitem{ni2022rtsnet}
------, ``{RTSNet}: Learning to smooth in partially known state-space models,'' \emph{{IEEE} Trans. Signal Process.}, vol.~71, pp. 4441--4456, 2023.

\bibitem{choi2023split}
G.~Choi \emph{et~al.}, ``Split-{KalmanNet}: A robust model-based deep learning approach for state estimation,'' \emph{{IEEE} Trans. Veh. Technol.}, vol.~72, no.~9, pp. 12\,326--12\,331, 2023.

\bibitem{buchnik2023latent}
I.~Buchnik \emph{et~al.}, ``Latent-{K}alman{N}et: Learned {K}alman filtering for tracking from high-dimensional signals,'' \emph{{IEEE} Trans. Signal Process.}, vol.~72, pp. 352--367, 2023.

\bibitem{wang2024nonlinear}
J.~Wang, X.~Geng, and J.~Xu, ``Nonlinear {K}alman filtering based on self-attention mechanism and lattice trajectory piecewise linear approximation,'' \emph{arXiv preprint arXiv:2404.03915}, 2024.

\bibitem{revach2022unsupervised}
G.~Revach, N.~Shlezinger, T.~Locher, X.~Ni, R.~J. van Sloun, and Y.~C. Eldar, ``Unsupervised learned {K}alman filtering,'' in \emph{European Signal Processing Conference (EUSIPCO)}.\hskip 1em plus 0.5em minus 0.4em\relax IEEE, 2022, pp. 1571--1575.

\bibitem{zhang2026change}
W.~Zhang, X.~Ni, N.~Shlezinger, and Z.~Wang, ``Change-aware self-adaptive {AI}-aided {K}alman filters with neural change point detection,'' \emph{arXiv preprint arXiv:2026.13387}, 2026.

\bibitem{chen2025maml}
S.~Chen \emph{et~al.}, ``{MAML-KalmanNet}: A neural network-assisted {K}alman filter based on model-agnostic meta-learning,'' \emph{{IEEE} Trans. Signal Process.}, vol.~73, pp. 988--1003, 2025.

\bibitem{ni2024adaptive}
X.~Ni, G.~Revach, and N.~Shlezinger, ``Adaptive {K}alman{N}et: Data-driven {K}alman filter with fast adaptation,'' in \emph{Proc. IEEE ICASSP}, 2024.

\bibitem{shlezinger2022discriminative}
N.~Shlezinger and T.~Routtenberg, ``Discriminative and generative learning for linear estimation of random signals,'' \emph{{IEEE} Signal Process. Mag.}, vol.~40, no.~6, pp. 75--82, 2023.

\bibitem{Dempster1977}
A.~P. Dempster, N.~M. Laird, and D.~B. Rubin, ``Maximum likelihood from incomplete data via the {EM} algorithm,'' \emph{Journal of the Royal Statistical Society: Series B (Methodological)}, vol.~39, no.~1, pp. 1--38, 1977.

\bibitem{cho2014learning}
K.~Cho, B.~van Merri{\"e}nboer, C.~Gulcehre, D.~Bahdanau, F.~Bougares, H.~Schwenk, and Y.~Bengio, ``Learning phrase representations using {RNN} encoder{--}decoder for statistical machine translation,'' in \emph{Proceedings of the 2014 Conference on Empirical Methods in Natural Language Processing ({EMNLP})}, 2014, pp. 1724--1734.

\bibitem{lorenz1963deterministic}
E.~N. Lorenz, ``Deterministic nonperiodic flow,'' \emph{Journal of the Atmospheric Sciences}, vol.~20, no.~2, pp. 130--141, 1963.

\bibitem{GannotDvorkind2006}
S.~Gannot and T.~G. Dvorkind, ``Microphone array speaker localizers using spatial-temporal information,'' \emph{EURASIP Journal on Applied Signal Processing}, vol. 2006, pp. 1--17, 2006.

\end{thebibliography}

\end{document}